\documentclass{ws-rv961x669}
\usepackage{ws-rv-van}     
\usepackage{ws-rv-thm}     
\usepackage{subfigure}     
\makeindex

\begin{document}

\chapter{
Exploring Quantum Hall Physics at Ultra-Low Temperatures and at High Pressures}


\author[F]{G\'abor A. Cs\'athy}

\address{Department of Physics and Astronomy, Purdue University \\
525 Northwestern Avenue, West Lafayette, IN 47907 \\
gcsathy@purdue.edu}

\begin{abstract}

The use of ultra-low temperature cooling and of high hydrostatic pressure techniques
has significantly expanded our understanding of the  
 two-dimensional electron gas confined to GaAs/AlGaAs structures.
This chapter reviews a selected set of experiments employing 
these specialized techniques in the study of the
fractional quantum Hall states and of the charged ordered phases, such as
the reentrant integer quantum Hall states and the quantum Hall nematic.
Topics discussed include a successful cooling technique used,
novel odd denominator fractional quantum Hall states, 
new transport results on even denominator fractional quantum Hall states and on
reentrant integer quantum Hall states, 
and phase transitions observed in half-filled Landau levels.

\end{abstract}


\body


\section{Introduction}
\label{gc-intro}

The two-dimensional electron gas (2DEG) is one of the richest model systems in condensed matter physics. 
Indeed, an impressive number of new phenomena were discovered in this system and several novel
theoretical concepts were introduced to explain these phenomena. The integer \cite{klitz}
and fractional quantum Hall effect \cite{tsui1} are among the most important discoveries in the 2DEG
and work on these Hall effects precipitated ideas on emergent quasiparticles \cite{laughlin,Jain,Halperin,mr,rr1}
and on topological concepts in condensed matter\cite{wenBook,steve}.

A large number of fractional quantum Hall states (FQHSs), especially the ones forming in the lowest Landau level,
are well understood\cite{tsui2,jainBook}. Their properties are accounted for by Laughlin's wavefunction\cite{laughlin} and Jain's
theory of composite fermions \cite{Jain,Halperin}. There are, however, a handful of fractional quantum Hall states
which are thought to harbor more intricate topological order. These states, sometimes referred to as
exotic fractional quantum Hall states\cite{jain-ar}, remain in the focus of current interest.

Historically, the two most studied 2DEGs were confined to MOSFETs 
and GaAs/AlGaAs heterostructures\cite{loren1}. 
In addition to these two examples, 2DEGs are supported by various
material hosts. Examples are AlAs/GaAlAs\cite{semi1}, CdTe/CdMgTe\cite{semi2}, 
 Si/SiGe\cite{semi3}, Ge/SiGe\cite{semi4,semi44}, ZnO/MgZnO\cite{semi5},
 hydrogen passivated Si surface\cite{semi6}, and electrons on the surface of superfluid Helium\cite{helium}.
The study of the 2DEG enjoyed a resurgence of interest with pioneering work on graphene\cite{geim} and 
other layered materials, such as transition metal dichalcogenides\cite{tmd} and black phosphorus\cite{bp}.
Work on the totality of 2DEGs highlighted some of the universal, 
host-independent physics. Examples are the formation of Landau levels in the integer quantum Hall regime
and of composite fermions
in the fractional quantum Hall regime. In addition, each of these hosts enriched the physics of the 2DEG.
New physics resulted from 2DEGs with novel internal degrees of freedom, such as the valley and the 
pseudospin quantum number, and also from 2DEGs possessing an inherent anisotropy.
Furthermore, the generation of Moir\'e lattices in layered van der Waals structures\cite{moire}
led to Hofstadter physics\cite{hofst} and, most recently, to magic angle superconductivity\cite{magic}.

\begin{figure}[t]
\centerline{\includegraphics[width=7cm]{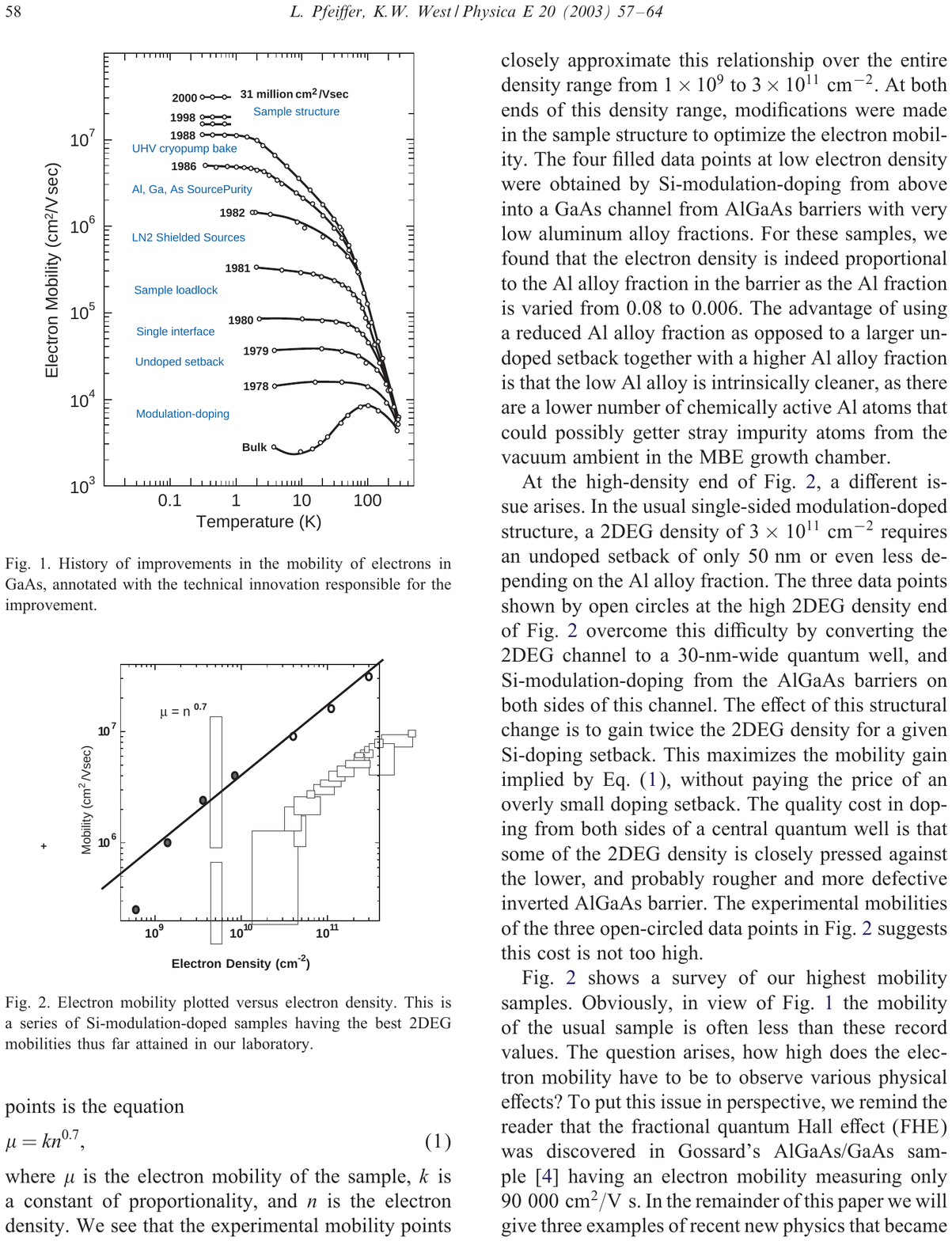}}
\caption{ Milestones in the evolution of the quality of 2DEGs in the GaAs/AlGaAs system, as measured
by the electron mobility. Adapted from Ref.\refcite{loren1}.}
\label{FigC-1}
\end{figure}

Even though 2DEGs confined to GaAs/AlGaAs heterostructures have been studied for more 
than three decades, this system continues to play a privileged role. A large number of phases 
were first seen in GaAs/AlGaAs. This is because numerous innovations in the 
Molecular Beam Epitaxy (MBE) growth technique culminated in 2DEGs of record 
high mobilities approaching $4 \times 10^7$~cm$^2$/Vs or record 
long mean free paths in excess of $0.3$~mm, when measured at temperatures below 1 Kelvin \cite{loren1}.
Such high carrier mobilities are possible because of the exceedingly low defect levels 
and also because of innovative sample structures. Growth efforts to further improve this material system
continue\cite{gaas1,gaas2,gaas3,watson15,gaas4,gaas5,gaas6}.
Historical milestones in the evolution of the GaAs MBE technology can be seen in Fig.~\ref{FigC-1}.

One area in which 2DEGs in GaAs/AlGaAs excel when compared to those in other hosts is
the support of both topological and traditional Landau phases. 
In contrast to topological phases,
traditional Landau phases may be characterized by an order parameter.
Charge ordered phases of the 2DEG are examples of such traditional Landau phases. The most well-known example 
of charge ordered phase is the Wigner crystal\cite{Wigner}. 
However, high quality 2DEGs allow for a more intricate charge ordering \cite{fogler,moessner,fradkin}.
The phase at half-filled Landau levels with a strong resistance anisotropy is commonly associated with
the electronic nematic, whereas the so called reentrant integer quantum Hall states (RIQHSs) are thought to
be identical to the electronic bubble phases \cite{lilly99,du99,fogler,moessner,fradkin}. 
Recently an increasing amount of attention
is lavished on the study of these charge ordered phases.


The 2DEG was probed with various ingenious techniques. However, electric transport played a special role among these
techniques as it historically was used to reveal new electronic phases. 
Over the last decade or so, transport experiments performed at ultra-low temperature 
have been especially fertile in providing new insight into the physics of the 2DEG. 
In the following we present a personal view on some of these experiments, focusing mainly
on the second Landau level of the 2DEG confined to  GaAs/AlGaAs hosts. 
In \sref{cs-sec2} the reader will find experimental details of the ultra-low temperature
cooling technique based on the He-3 immersion cell.
\Sref{cs-sec3} discusses recently discovered FQHSs, all of which are at odd denominators,
followed by an examination of possible origins of these states.
In \sref{cs-sec4} recent results of transport experiments on the
even denominator FQHSs are discussed. Topics include 
phase transitions at the Landau level filling factor $\nu=5/2$, 
a discussion of the energy gap of the FQHS at this filling factor, 
and its behavior in the presence of short-range alloy disorder.
\Sref{cs-sec5} contains results on RIQHSs in the second Landau level, 
such as magnetoresistive fingerprints of these states in high mobility samples, 
a discussion of the precursors of these states, and a summary of new, developing RIQHSs.
Finally, the pressure-induced phase transition at $\nu=5/2$ from a FQHS
to the quantum Hall nematic is discussed in \sref{cs-sec6}.

\section{Cooling electrons below 10 mK and sample state preparation}
\label{cs-sec2}

Lowering the electronic temperature typically allows for the resolution of states of reduced energy scales. 
In other words, unless the energy spectrum is already disorder dominated,
a lower temperature may reveal increasingly more fragile electronic ground states. 
Reducing the electron temperature, however, is not a trivial task. 
While modern dilution refrigerators routinely generate mixing chamber temperatures below $10$~mK, 
similarly low electronic temperatures in semiconductor nanostructures are often difficult to achieve. 
This is because the combination of reduced electron-phonon coupling 
at milliKelvin temperatures and of minute amounts of uncontrolled radiofrequency power traveling 
on the measurement wires; under such conditions the electron temperature in a transport setup is 
often higher than that of the phonons and also of the coldest spot of the refrigerator.

\begin{figure}[t]
\centerline{
  \minifigure[Schematic of a He-3 immersion cell for ultra-low temperature transport measurements of 2DEGs.
     Adapted from Ref.\refcite{setup}.]
     {\includegraphics[width=2.1in]{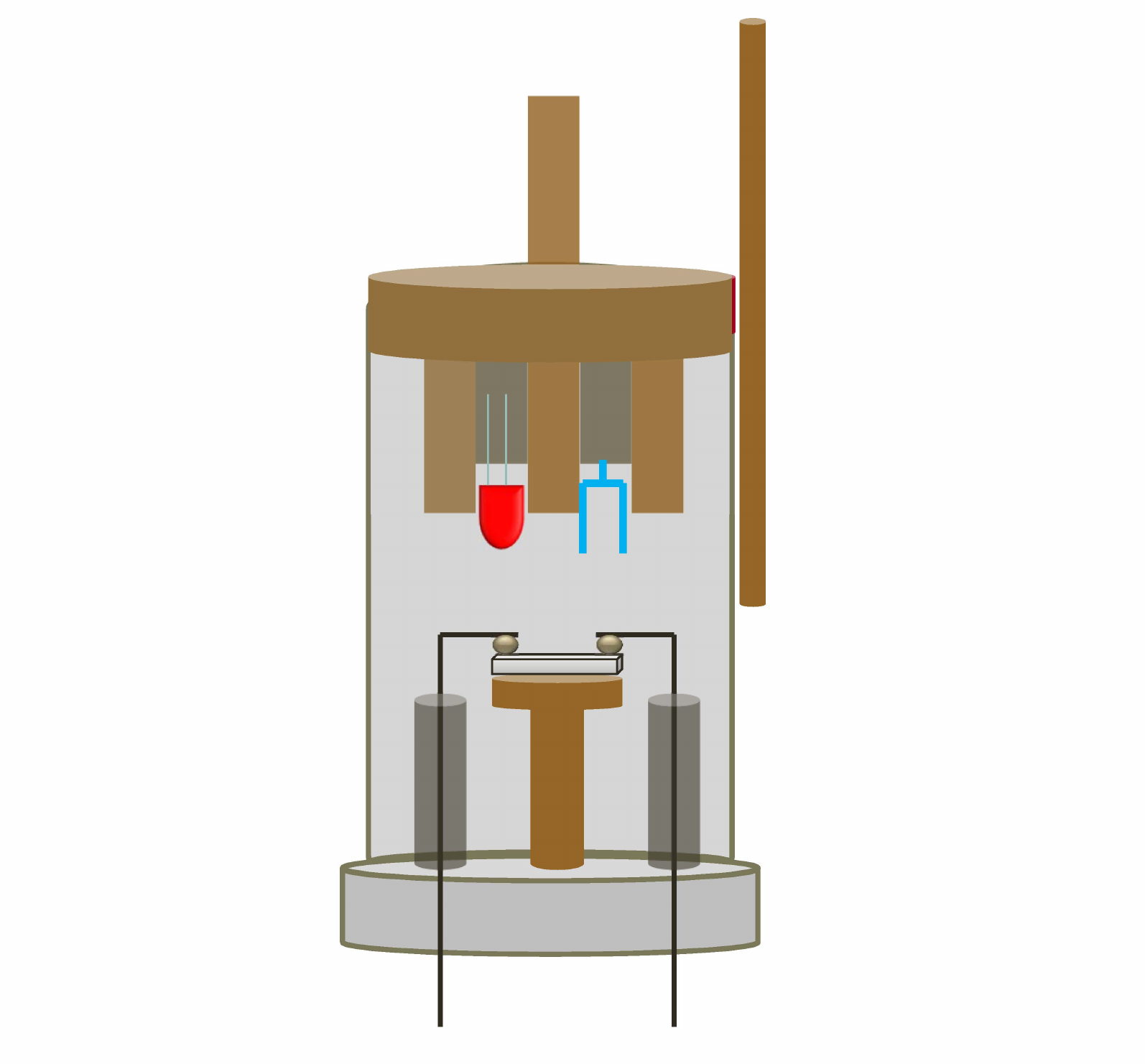}\label{FigC-2}}
 \hspace*{0pt}
  \minifigure[A quartz tuning fork and the
     temperature dependence of its response when
     immersed into a He-3 bath. Adapted from Ref.\refcite{setup}.]
     {\includegraphics[width=2.4in]{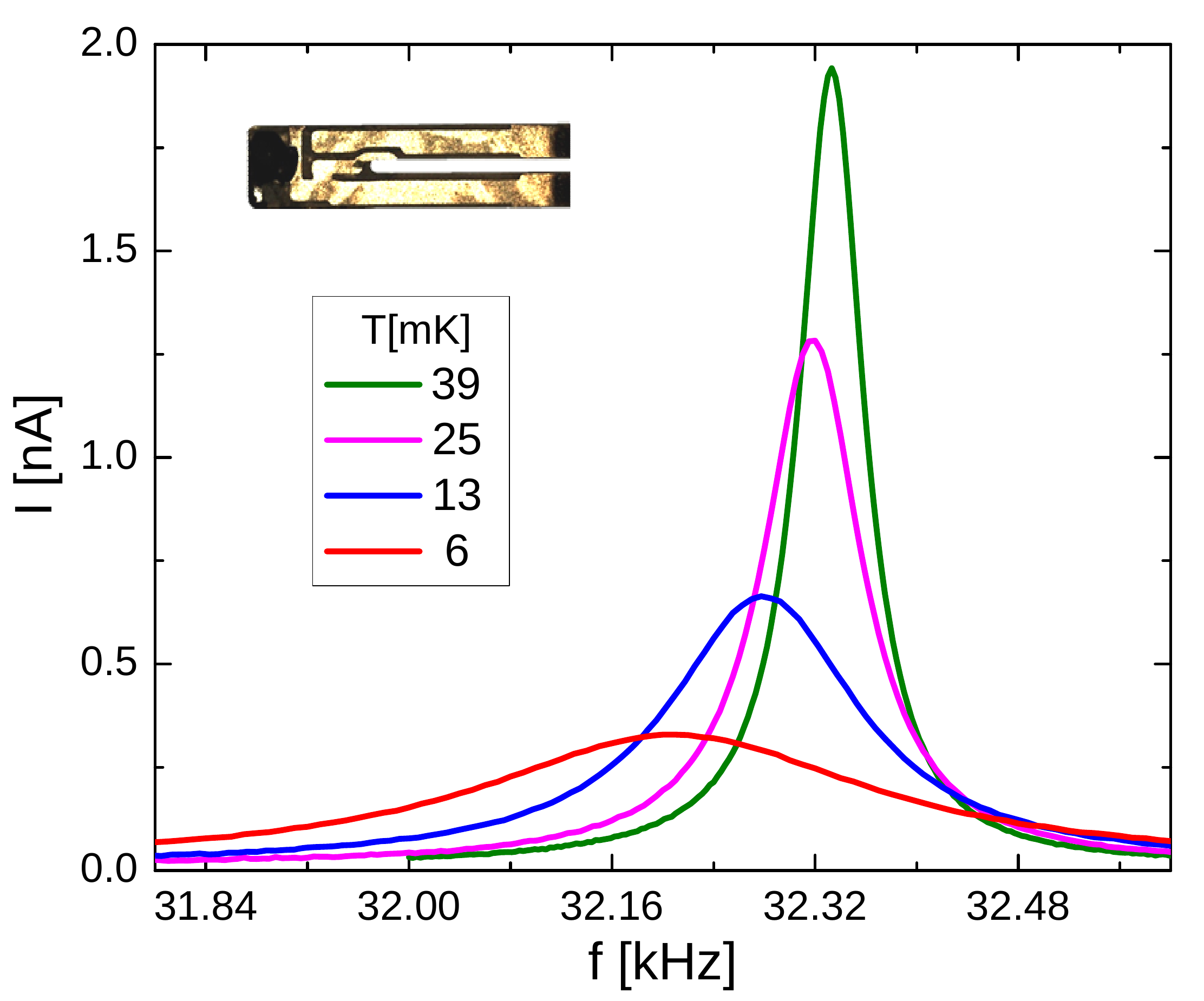}\label{FigC-3}}
}
\end{figure}

In this section we describe a 
successful electron thermalization setup based on a He-3 immersion cell.
Such an immersion cell was first built by Jian-Sheng
Xia and coauthors at the MicroKelvin Facility of the National High Magnetic Field Laboratory in Gainesville\cite{xia00}. 
In this setup each ohmic contact of an electrically conductive sample, 
such as a 2DEG, is soldered onto individual wire heatsinks that consist of a silver wire surrounded by silver sinter. 
Constrained by the geometry of the superconductive magnet bore, one may achieve 
a surface area of the order of a m$^2$ for the sinter of each wire heatsink.
In order to take advantage of such a large surface 
area for thermalization purposes, the 2DEG and the wire heatsinks are immersed into liquid He-3 that ensures 
both thermalization and electrical isolation. 
This thermalization setup was used to examine the quantization at 
$\nu=5/2$\cite{pan99}, to discover the $\nu=12/5=2+2/5$ 
FQHS\cite{xia04}, to study of metallic behavior in a hole gas\cite{huang}, and
to understand the plateau-to-plateau transition in the integer quantum Hall regime\cite{wanli}. 
A schematic of a  He-3 immersion cell is shown in Fig.~\ref{FigC-2}.
Xia et al. have also built a He-3 immersion cell with a hydraulically driven rotator stage for tilted field measurements\cite{xia02}.
Since cooling electronic systems to ultra-low temperatures is increasingly important 
for the study of fragile electronic order, besides the immersion cell technology there are 
also efforts to develop alternative cooling techniques\cite{cool-1,cool-2,cool-2b,cool-3}, some of which are designed
on cryogen free platforms.

For our He-3 immersion cell we adopted the design from Ref.\refcite{xia00}. In contrast to the setup in Gainesville, 
instead of using a dilution refrigerator equipped with a nuclear demagnetization stage, we have attached our 
immersion cell to a modified dilution refrigerator with a $5$~mK base temperature. Construction details
can be found in Ref.\refcite{setup}.
Microwave filters installed on measurement wires are a critical part of the setup. 
We used a three-stage filter on each wire. First, a set of capacitive filters mounted on the top of the refrigerator, 
as part of a D-sub connector, is used at room temperature. Second, on their way to the sample, the 
measurement wires are well heatsunk at each stage of the refrigerator and passed through silver epoxy
embedded along about the one foot length of tail connecting the immersion cell
to the mixing chamber.
Finally, through the skin effect, silver sinters of each wire heatsink mounted within the immersion cell 
will also efficiently dissipate microwaves\cite{setup}. Additional low-pass RC filters with a cut-off frequency
of $50$~kHz and mounted on the still did not make a difference in
electron thermalization, therefore they were later removed. 

Parallel with the development of the cooling technology, there was also a flurry of activities in
thermometry. Examples are thermometers based on resistive elements\cite{thermo-0}, 
on Johnson noise measurements\cite{thermo-1,thermo-2,thermo-3}, 
and on measurement of the tunneling conductance\cite{thermo-3,thermo-4,thermo-5,thermo-6,thermo-7}. 
The extreme environment of temperatures below $10$~mK and strong magnetic fields 
limit the choice of thermometers.
The widely used RuO resistive sensors become unreliable for thermometry below about $20$~mK, especially in strong fields. 
Paramagnetic susceptibility thermometers are not suitable for operation in strong magnetic fields. 
While He-3 melting curve thermometers could be used, we were deterred by the additional effort needed for 
handling the He-3 at high pressures. However, we already had liquid He-3 in the immersion cell for thermalization purposes, 
albeit the He-3 was not at the high pressures need for operation at the liquid-solid phase boundary required
in a melting curve thermometer. Under such conditions the
viscosity of the He-3 liquid provides a convenient way for temperature monitoring from
about 100~mK down to the superfluid onset temperature (not within the reach of our instrument). 
Since viscosity is independent of the magnetic field, it is ideally suited for the demanding 
low temperature and strong magnetic field environment of measurements in the quantum Hall regime. 
We opted for a quartz tuning fork based viscometer\cite{setup}. An example of a quartz tuning fork and
the temperature dependence of the response curve of the tuning fork immersed into liquid He-3 are shown
Fig.~\ref{FigC-3}.

Finally, the electronic state of 2DEGs confined to GaAs/AlGaAs is often prepared by a brief low temperature
illumination using either a red light emitting diode facing the 2DEG\cite{illum-1,illum-2} or light 
guided towards the sample using fiber optics\cite{folk}. One effect of such illumination is the increase of the electron density
that is desirable for robust ground states. In addition, illumination may also improve the homogeneity of the
electron gas.

\section{Recently discovered fractional quantum Hall states}
\label{cs-sec3}

In this section we discuss new FQHSs discovered in the 2DEG in the GaAs/AlGaAs system.
Two of these, the FQHSs at $\nu=2+6/13$  and at $\nu=3+1/3$, were seen in the 
region of the second Landau level, whereas the FQHSs at $\nu=4/11$ and $\nu=5/13$
develop in the lowest Landau level. These FQHSs are fragile, hence the use of the immersion cell technology
played a central role in their discovery. We show that an analysis of the energy gaps
reveals valuable insight into the nature of these FQHSs.


Ground states with a second Landau level character are the strongest 
in 2DEGs with the optimal  electron density. Indeed, 
as a rule of thumb, the energy gap of a given FQHS increases with the value of the magnetic field at which it forms. Therefore 
in order to maximize the energy gap of a FQHS, one must use a 2DEG with the largest possible density. However, because
the finite thickness of the electronic wavefunction in the direction perpendicular to the plane of the 2DEG,
past a critical density the Fermi level moves to the lowest Landau level of the
second electric subband. As shown by Shayegan and 
coauthors\cite{shay10,shay11} and also supported by theory\cite{papi}, such a
population of the second electrical subband will have a strong influence on 
the ground states: the orbital part of the single particle wavefunction 
 changes from a second Landau level character to
a lowest Landau level character. The most dramatic consequence
of populating the second electrical subband is the rapid collapse of the
FQHS at $\nu=5/2$\cite{shay10,shay11} . 
The optimal density is therefore the largest density at which the Fermi level falls into the
second Landau level and at which ground states retain a second Landau level character.
For a quantum well of a
$30$~nm width, the optimal density is about $3.0 \times 10^{11}$~cm$^{-2}$.

\begin{figure}[t]
\centerline{\includegraphics[width=1\columnwidth]{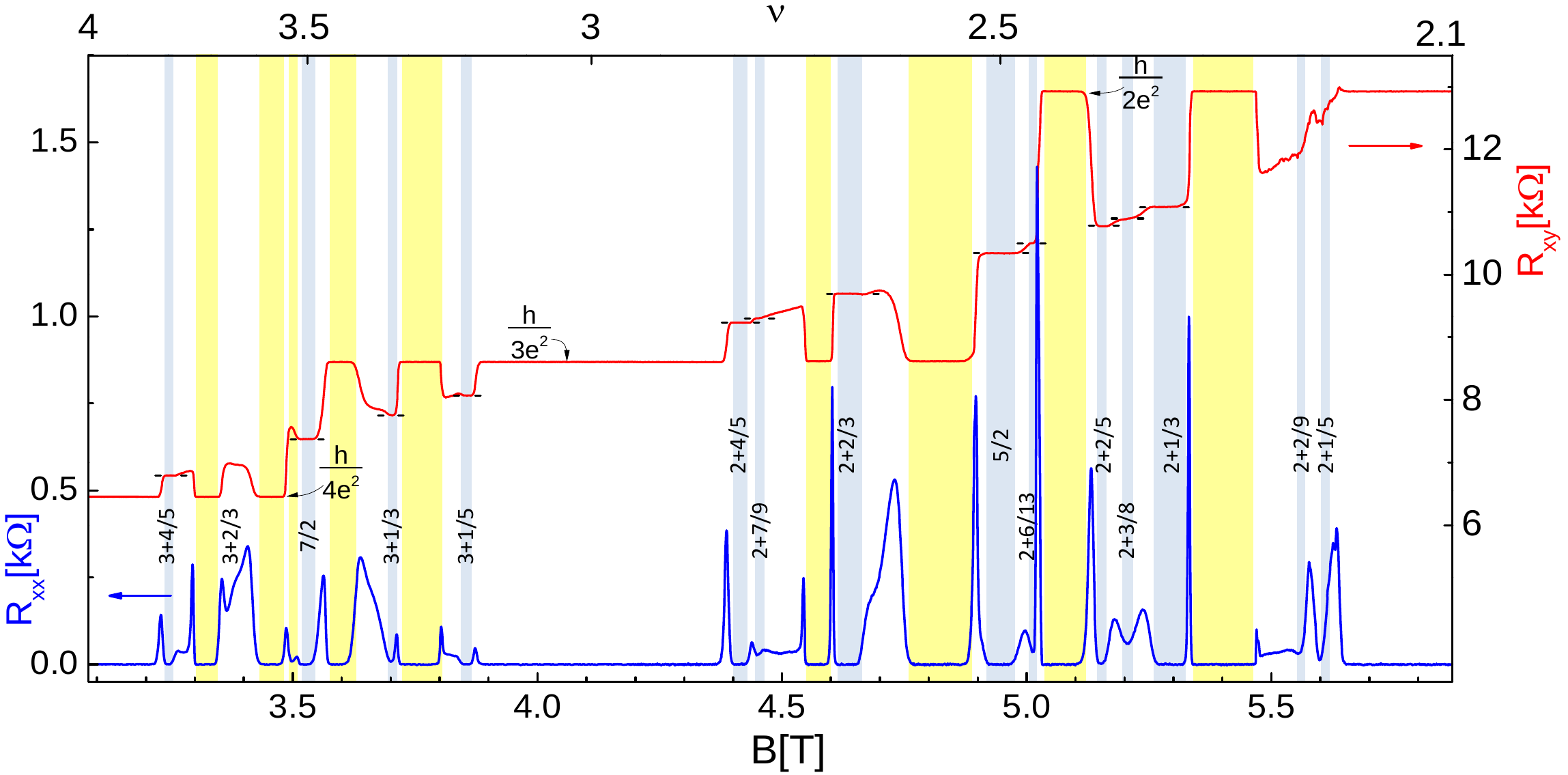}}
\caption{A particularly rich magnetotransport trace measured at 6.9~mK in the second Landau level. 
   Both spin branches are shown.
   Blue shades mark fractional quantum Hall states (FQHSs) at the labeled filling factors, 
   whereas yellow shades show various reentrant integer quantum Hall states (RIQHSs). Adapted from Ref.\refcite{ethan15}.}
\label{FigC-4}
\end{figure}

Fig.~\ref{FigC-4} shows magnetoresistance traces with a particularly rich structure
over the the full width of the second Landau level, i.e. both the lower and upper
spin branches, in another 2DEG near its optimal density\cite{kumar10,ethan15}.  
FQHSs are shaded in blue, while RIQHSs in yellow.
At first blush, magnetoresistance in the lower spin branch of the second Landau level
shown in Fig.~\ref{FigC-4} is similar to that in Ref.\refcite{xia04}. However, a more careful examination reveals
a notable difference in the sharpness of several magnetoresistance peaks.
For example, the width at half height of the peak in $R_{xx}$ shown in Fig.~\ref{FigC-4}
near $B=5.02$~T is $6.6$~mT, while of that near $B=5.33$~T is merely $4.2$~mT.
In contrast, the corresponding peaks in Ref.\refcite{xia04} are considerable wider.
The differences in the two traces are surprising, given that samples and measurement 
conditions were comparable: the electron densities, mobilities, and the fridge temperatures were
$n=3.1 \times 10^{11}$~cm$^{-2}$, $\mu= 31 \times 10^6$~cm$^2$/Vs, $T=9$~mK in Ref.\refcite{xia04}, and
$n=3.0 \times 10^{11}$~cm$^{-2}$, $\mu= 32 \times 10^6$~cm$^2$/Vs, $T=6.9$~mK in 
Refs.\refcite{kumar10,ethan15}.
The difference in the sharpness of the resistance peaks is likely attributed to a difference
in sample homogeneity. Indeed, in the presence of a density variation, the magnetoresistance
will be a convolution of magnetoresistances corresponding to the range of electron densities 
and therefore a density inhomogeneity will lead to a broadening of sharp resistive features.
Density inhomogeneity in high quality GaAs/AlGaAs samples
may be estimated from the widths of sharp magnetoresistance peaks\cite{pan05}, from an analysis of 
quantum lifetime measurements\cite{qian17}, and 
it is also accessible with the micro-photoluminescence technique 
on lengthscales larger than the laser spotsize\cite{mPL1,mPL2}.
We believe that the improved sample homogeneity results from a combination of improved
sample growth and sample illumination techniques.

\subsection{$\nu=2+6/13$ and $\nu=2+2/5$ fractional quantum Hall states}
\label{cs-sec3p1}

A particularly interesting region within the second Landau level
is that of $2+1/3 < \nu < 2+2/3$, which was conjectured to host FQHSs with unusual topological 
order\cite{wojs09}. Several FQHSs in this region may have exotic fractional
correlations\cite{mr,rr1,gaffn,bishara,bonderson,bernevig,levin,bipart,tripart}.
The FQHSs at the two endpoints of this range may be either Laughlin states or 
states radically different from it\cite{ambum88,scarola01,toke05,papic09,ajit13,johi14,peters15,joli17,park17}.
Until 2004, FQHSs in this region were detected, using a He-3 immersion cell, 
at $\nu=5/2$, $\nu=2+2/5$, and $\nu=2+3/8$ \cite{xia04}. An account of FQHSs in GaAs/AlGaAs, 
including of the ones in the second Landau level, was published in 2008\cite{pan08}.

Another immersion cell experiment confirmed a fully quantized $\nu=2+2/5$ FQHS, 
with an energy gap of $80$~mK \cite{kumar10} in the $2+1/3 < \nu < 2+2/3$ range.
It also yielded a new FQHS in this range at $\nu=2+6/13$\cite{kumar10}.
As seen in Fig.\ref{FigC-4}, this FQHS is located between the $\nu=5/2$ FQHS and a RIQHS.
Since its first observation, the $\nu=2+6/13$ FQHS has been seen in an increasing number of 
experiments\cite{watson15,zhang12,deng12,qianNat}.

\begin{figure}[t]
\centerline{
  \minifigure[Details of the temperature dependence of the magnetoresistance near the RIQHS $R2b$.
  Vertical arrows mark the precursor of the RIQHS $R2b$. Adapted from Ref.\refcite{shingla}.]
     {\includegraphics[width=2.6in]{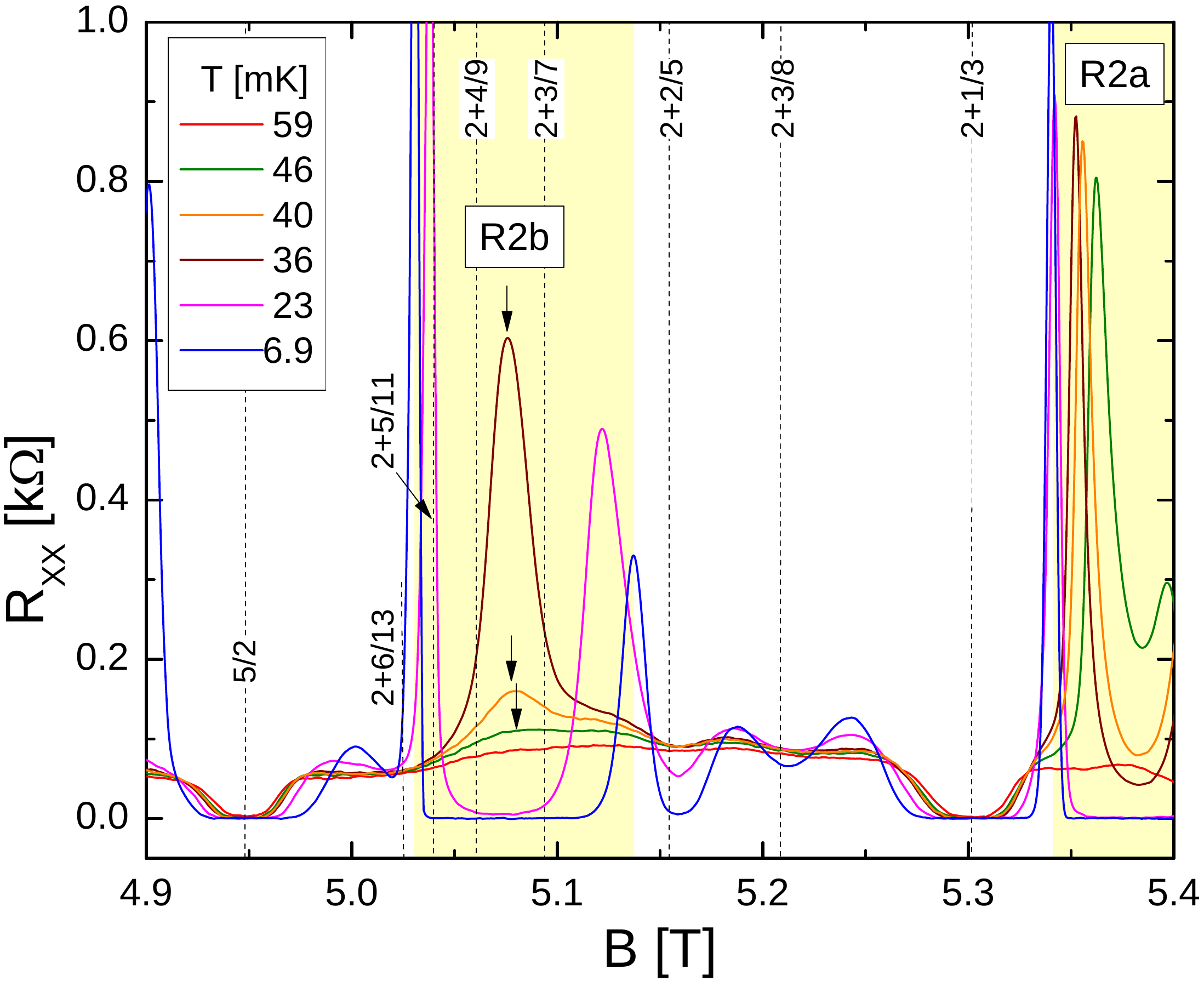}\label{FigC-5}}
  \hspace*{0pt}
  \minifigure[Energy gaps of FQHSs in the $2+1/3 < \nu < 2+2/3$ range versus $|B_{eff}|$. 
  Adapted from Ref.\refcite{kumar10}.]
     {\includegraphics[width=1.87in]{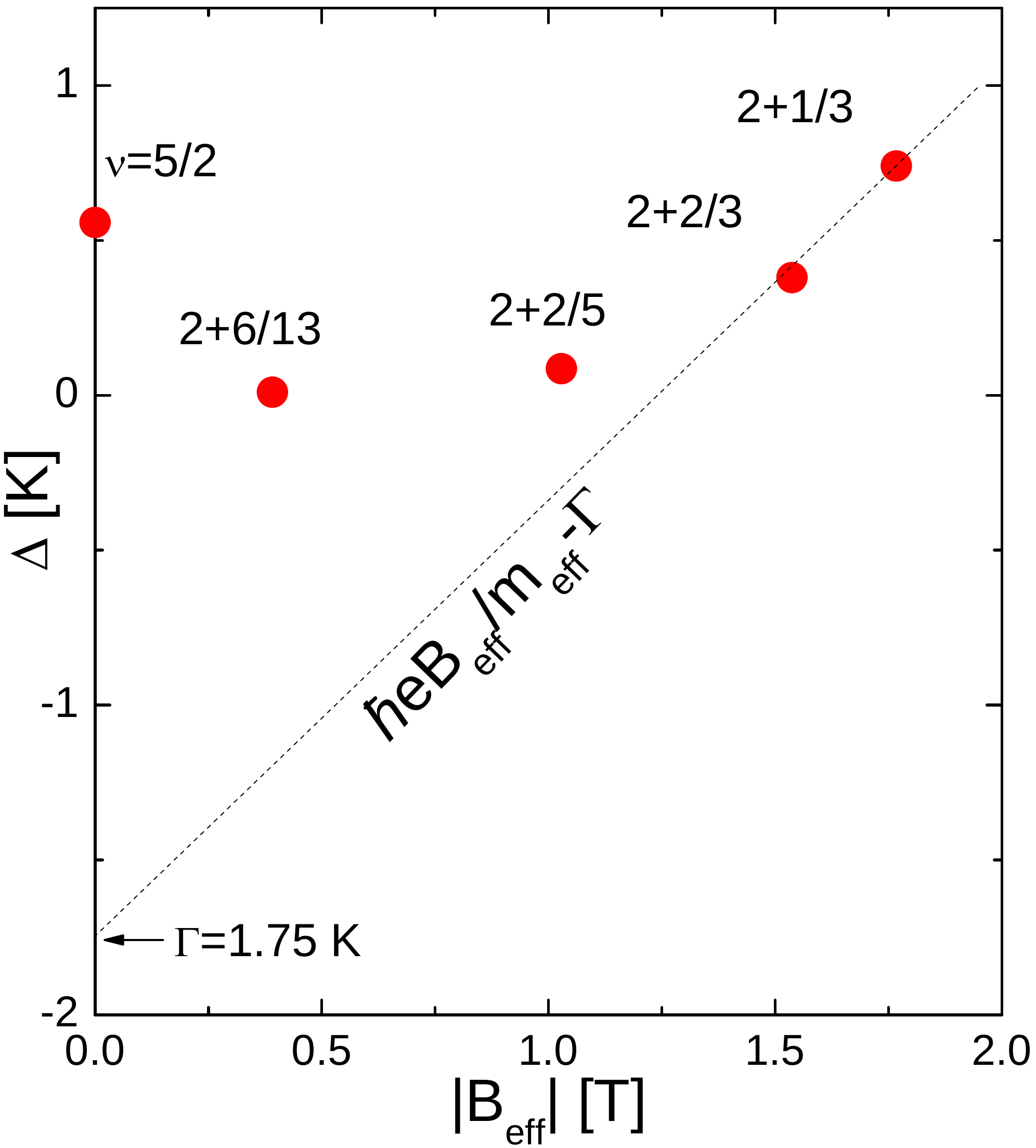}\label{FigC-6}}
}
\end{figure}

The observation of a FQHS at $\nu=2+6/13$ was a surprise. Indeed,
Jain's model of non-interacting composite fermions\cite{Jain} offers a concise and elegant framework for the understanding 
of FQHSs developing at filling facors of the form  $p/(2p \pm 1)$, with $p=1,2,3,...$.
One may notice that the subset of $\nu=2+1/3$, $2+2/5$, and $2+6/13$ filling factors, at which
there are FQHSs in the second Landau level, also belongs to the Jain sequence
of the form $2+p/(2p+1)$, with $p=1,2,$ and $6$, respectively. 
However, other FQHSs at intermediate values of $p=3,4,$ and $5$ are not present in Fig.~\ref{FigC-4}.
As shown in Fig.~\ref{FigC-5}, the missing FQHSs are absent not only at the lowest temperatures, 
but for a range of temperatures\cite{shingla}. 
Because of these absent FQHSs with $p=3,4,$ and $5$, it was unexpected to
see a FQHS of an unusually high order $p=6$ at $\nu=2+6/13$. 
We note that even though in other experiments local minima in $R_{xx}$ were seen at $\nu=2+4/9$ 
in Refs.\refcite{gaas3,choi08} and at $\nu = 2 + 3/7$ in Ref.\refcite{choi08}, in the the absence of a
Hall plateau the identification of a FQHS at these filling factors is inconclusive \cite{gaas3,choi08}.

At this time it is not known why the $\nu=2+3/7, 2+4/9,$ and $2+5/11$ FQHSs are not present in Fig.~\ref{FigC-5}.
One explanation for the the missing FQHSs at $\nu=2+3/7, 2+4/9,$ and $2+5/11$ may be offered by
a competition with the RIQHS labeled $R2b$. 
If the energies of the competing RIQHS and FQHSs are comparable, one may expect incipient
FQHSs near the onset temperature of the RIQHS. However, magnetotransport did not exhibit
such features near the onset of the RIQHS $R2b$\cite{shingla}. Indeed,
gentle curvatures of the magnetoresistance seen  in Fig.~\ref{FigC-5}
near $\nu=2+3/7$ and $2+4/9$ in traces measured at $T=59$, $46$, and $40$~mK
could not be assigned to incipient FQHSs; instead these features
have been associated with magnetoresistive fingerprints of the precursor of the RIQHS\cite{shingla}.
This suggests that the RIQHS labeled $R2b$ is considerably more stable than the FQHSs and therefore
it is strongly favored from an energetical point of view.
According to a second argument presented in the following,
if the FQHSs at $\nu=2+2/5$ and $2+6/13$ have topological order different from
that of the model of non-interaction composite fermions, then 
FQHSs at the intermediate filling factors $\nu=2+3/7, 2+4/9,$ and $2+5/11$
do not have to necessarily exist.

The development of FQHSs at filling factors belonging to the Jain sequence $2+p/(2p+1)$
does not necessarily mean that these FQHSs can be described by the model of 
non-interacting composite fermions.
Indeed, there are ideas proposed according to which in the presence of strong residual
interactions between the composite fermions, the nature of a FQHS may be different from
that predicted by the model of non-interacting composite fermions. An examination of
the energy gaps of these FQHSs provided early insight in this regard\cite{kumar10}. 

In the following we discuss energy gaps of the odd denominator FQHSs of the second Landau level.
We will use a widely used phenomenological model\cite{dis1,dis2}. 
The purpose of this model is to bridge the difference between energy gaps obtained from numerical
experiments and those from measurements. The former are calculated in simulations that do not include
any disorder, while the latter are extracted from measured data in real 2DEGs that necessarily
have some disorder present.
According to this simple model, the measured energy gap $\Delta$ is
reduced from the expected theoretical value in the limit of no disorder, also called the intrinsic gap
$\Delta^{int}$, by an amount due to disorder broadening $\Gamma$:
\begin{equation}
\Delta=\Delta^{int}-\Gamma . 
\label{eq1}
\end{equation}
The intrinsic gap $\Delta^{int}$ contains effects of the finite width of the wavefunction 
in the direction perpendicular to the plane of the 2DEG and of Landau level mixing.
Since disorder effects are not included into $\Delta^{int}$, 
it may therefore be directly compared to gaps from numerical experiments, 
such the ones from exact diagonalization. 
Eq.(\ref{eq1}) therefore offers a simple way to deal with the effects of the disorder on the energy gap, 
a task that remains challenging for the theory. The concept of disorder broadening
was used to understand early gap measurements of FQHSs\cite{dis1,dis2} and
also of the series of the FQHSs in the lowest Landau level\cite{duGap}.
In the latter work, the dependence of gaps on the effective magnetic field was found linear.
The slope of this linear dependence yields the cyclotron mass of the composite fermions\cite{duGap}. 
Furthermore, it was found that energy gaps extrapolate to a negative offset at
zero effective magnetic field\cite{duGap}. This negative offset was identified with 
a filling factor independent $\Gamma$, with values between $1$ and $2$~K. 

In the second Landau level the number of FQHSs is considerably less than that in the lowest Landau level.
Nonetheless, as shown in Fig.~\ref{FigC-6},
the simplest possible analysis involving a reflection of the FQHS at $\nu=2+2/3$ to positive effective magnetic fields
shows that a linear dependence of the gaps does no longer hold in the second Landau level\cite{kumar10}.
The complex dependence of the energy gaps on the effective magnetic field
 indicates that residual interactions of the composite fermions are significant in
the second Landau level and, as a result, at least some of the FQHSs may be beyond the model
of non-interacting composite fermions\cite{kumar10}.
If in addition the effective mass of composite fermions in the second Landau level
is assumed to be the same as of those forming in the lowest Landau level,
one finds that the energy gaps of the $\nu=2+1/3$ and $2+2/3$ FQHSs are consistent
with the model of free composite fermions when the disorder broadening is $\Gamma=1.75$~K\cite{kumar10}.
The expected energy gaps of FQHSs of free composite fermions
under these assumptions are shown by a dashed line in Fig.~\ref{FigC-6}.
The proximity of the measured gaps at $\nu=2+1/3$ and $2+2/3$ to the dashed line 
shown in Fig.~\ref{FigC-6} suggests that FQHSs at $\nu=2+1/3$ and $2+2/3$ may be of the Laughlin type.
Similar conclusions on the nature of the $\nu=2+1/3$ and $2+2/3$ FQHSs
were also drawn in studies of the neutral modes at $\nu=2+1/3$ and $2+2/3$
\cite{neutral}, edge-to-edge tunneling experiments at $\nu=2+2/3$\cite{tunn1},
and magnetoroton measurements at $\nu=2+1/3$\cite{pinczuk}.
Nonetheless, overlap calculations of the numerically obtained wave
function with the Laughlin state and other theory work suggests that
the FQHS at $\nu=2+1/3$ is not yet satisfactorily
understood\cite{ambum88,scarola01,toke05,papic09,ajit13,johi14,peters15,joli17,park17}.
The analysis presented above is the simplest possible one; an
alternative model for the energy gaps at odd denominators may be found in Ref.\refcite{manoGap}.

Under the assumptions described above, the measured energy gaps of both the $\nu=2+6/13$ and
$\nu=2+2/5$ FQHSs are significantly larger than the expected values of a model of non-interacting 
composite fermions \cite{kumar10}.
This discrepancy of the energy gaps, also shown in Fig.~\ref{FigC-6}, constituted an early 
experimental evidence
that the FQHSs at $\nu=2 +6/13$ and $2+2/5$ are not similar to
their lowest Landau level counterparts forming at $\nu=6/13$ and $2/5$, but they are likely
of exotic, perhaps non-Abelian nature. 
The argument according to which the interactions between the composite fermions
at $\nu=2+6/13$ are significant enough to change the topological order of
the FQHSs at this filling factor is not unreasonable, since similar
interactions at the nearby filling factor $\nu=5/2$ are known to drastically change
the ground state from a Fermi sea of composite fermions to a FQHS.
In fact the FQHSs at $\nu=2+6/13$ and at $\nu=5/2$ may be closely related.
Indeed, according to a recent proposal, FQHSs at $\nu=2+6/13$ and $\nu=5/2$ as well as
other FQHSs of the second Landau level
may be accounted for within the same description based on
parton wavefunctions \cite{parton1,parton2,parton3}.

The above analysis of the gaps has implications not only for the FQHS at $\nu=2+6/13$, but also 
for the FQHS at $\nu=2+2/5$. According to this analysis, the $\nu=2+2/5$ FQHS is
also an exotic FQHS\cite{kumar10}.
Proposals for the description of the $\nu=2+2/5$ FQHS include
the Read-Rezayi parafermion construction\cite{rr1}, the particle-hole conjugate of this state\cite{bishara},
the Bonderson-Slingerland state\cite{bonderson}, the Gaffnian\cite{gaffn}, 
the Levin-Halperin Abelian construction\cite{levin}, multipartite composite fermion states\cite{bipart,tripart},
and a parton state\cite{parton3}.
Numerical work favors the Read-Rezayi description \cite{rr1,rr2,zhu15,pakr16,papic17}, 
which however is in close competition with the Bondenson-Slingerland state\cite{bonders12} .
It appears that no single theory among the ones predicting exotic behavior at
$\nu=2+2/5$ and $2+6/13$ can account for FQHSs at
these two filling factors in a natural way and therefore one may surmise that
these two FQHSs have fundamentally different origins.

While at $\nu=2+6/13$ there is a visible FQHS, a FQHS at the particle-hole symmetry related filling factor
of $\nu=2+7/13$ is conspicuously missing\cite{kumar10}. A similar lack of particle-hole symmetry is also
observed at partial filling $2/5$. Indeed, as already discussed, the ground state at $\nu=2+2/5$ is
a FQHS, whereas at $\nu=2+3/5$ a FQHS does not form. 
The lack of particle-hole symmetry at $\nu=2+7/13$ and $\nu=2+3/5$  affects FQHSs 
differently than the RIQHSs \cite{shingla}.
It is interesting to note that a related particle-hole asymmetry was also observed in a high quality 
bilayer graphene\cite{gr1}. In this experiment a weak FQHS is seen at $\nu=2+7/13$, but not at $\nu=2+6/13$. 
However, in Ref.\refcite{gr1} a competition with a RIQHS cannot be invoked.
This result indicates that a particle-hole symmetry breaking effect, 
such as Landau level mixing, is likely at play and it favors one set of FQHSs, while suppresses
fractional correlations at the particle-hole symmetric filling factor.

Finally, we note that data discussed above are all in the single layer limit, i.e. when the second electrical subband
is not occupied. Recent work on the $\nu=2+2/5$ and $2+3/5$ FQHSs in a wide quantum well in
a GaAs/AlGaAs system in which the second subband is occupied suggests that these two 
FQHSs belong to the model of non-interacting composite fermions\cite{shay10}. Once
the orbital wavefunctions acquire a lowest Landau level character in samples with the second electrical
subband occupied, particle-hole symmetry of the FQHSs appears to be restored\cite{shay10}.

\subsection{$\nu=3+1/3$ fractional quantum Hall state}

The upper spin branch of the second Landau level, i.e. the $3 < \nu < 4$ region,
is at a lower magnetic fields and therefore hosts fewer FQHSs than the lower spin branch. 
Early work on this region established the FQHSs at $\nu=7/2$, $\nu=3+1/5$, and $\nu=3+4/5$
as well as four RIQHSs\cite{eisen02}.
The FQHS discovered at $\nu=3+1/3$ is the most recently seen ground state 
in this region\cite{ethan15}. A FQHS as at this filling factor is identified by a vanishing magnetoresistance and a Hall
resistance quantized to $h/(3+1/3)e^2$.

Even though in Fig.\ref{FigC-4}  a distinct minimum in the magnetoresistance 
is also observed at $\nu=3+2/3$, a FQHS at this filling factor is conspicuously missing\cite{ethan15}.
Indeed, an examination of the temperature dependence of the magnetoresistance revealed that 
the local minimum at $\nu=3+2/3$
does not follow the usual decreasing trend with a decreasing temperature. Therefore the opening
of an energy gap, a defining property of FQHSs, could not be established at $\nu=3+2/3$.
Furthermore, the Hall resistance at $\nu=3+2/3$ was not quantized, in fact its value
was not close to the expected value of $h/(3+2/3)e^2$ for a FQHS. Taken together, the
existence of a fractional quantum Hall ground state at $\nu=3+2/3$ so far could not be established\cite{ethan15}.

\begin{figure}[t]
\centerline{
  \minifigure[Energy gaps in the second Landau level at partial filling $1/3$ versus the Coulomb energy. 
     Data taken from Refs.\refcite{kumar10,ethan15}.]
     {\includegraphics[width=1.53in]{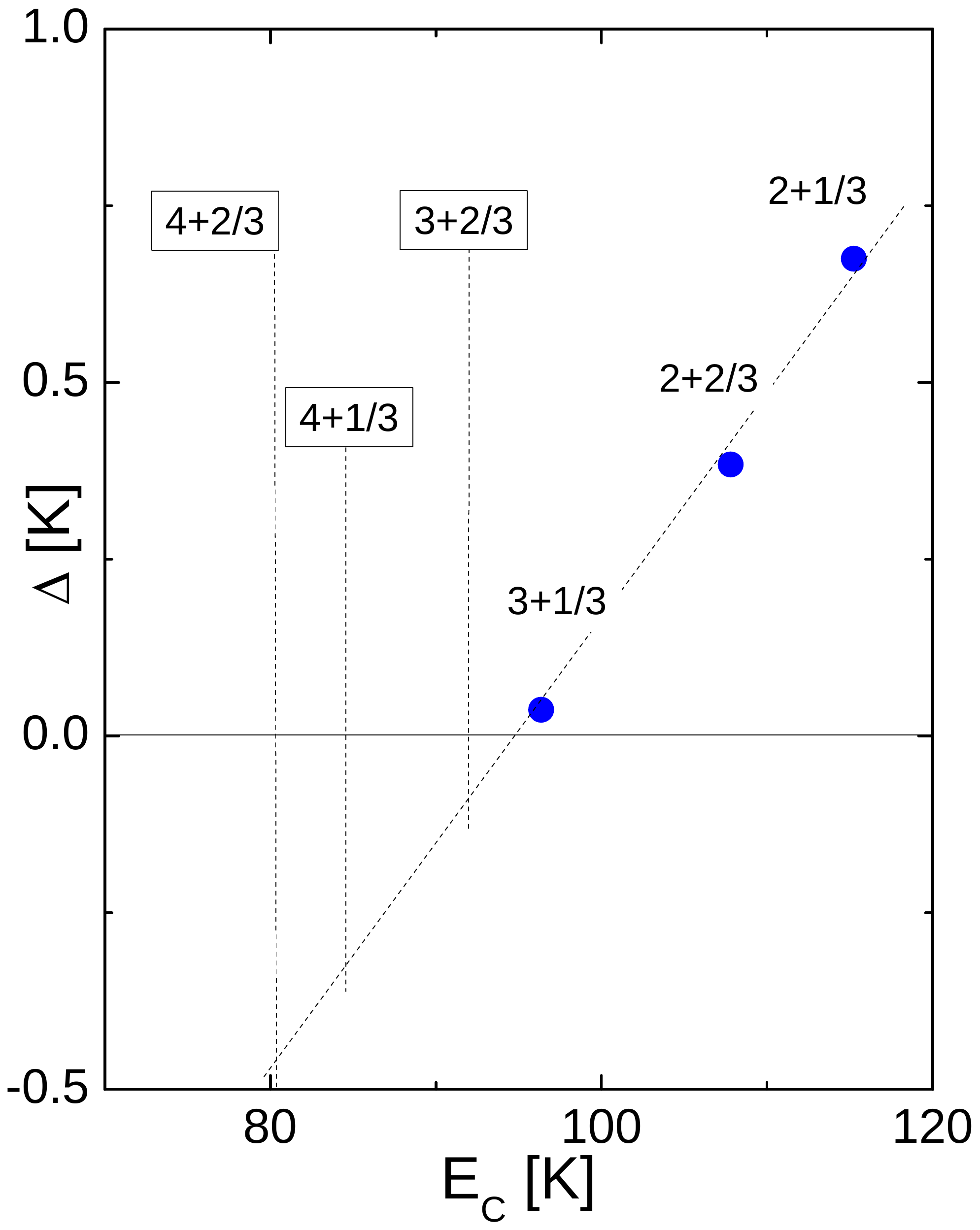}\label{FigC-7}}
  \hspace*{0pt}
  \minifigure[Magnetotransport in the second Landau level in a sample of low density. 
     Blue shades mark FQHSs at the labeled filling factors, whereas yellow shades show precursors of eight RIQHSs. 
     The lower spin branch data between $B=1.1$ and $1.7$~T is adapted from Ref.\refcite{nodar11}.]
     {\includegraphics[width=3in]{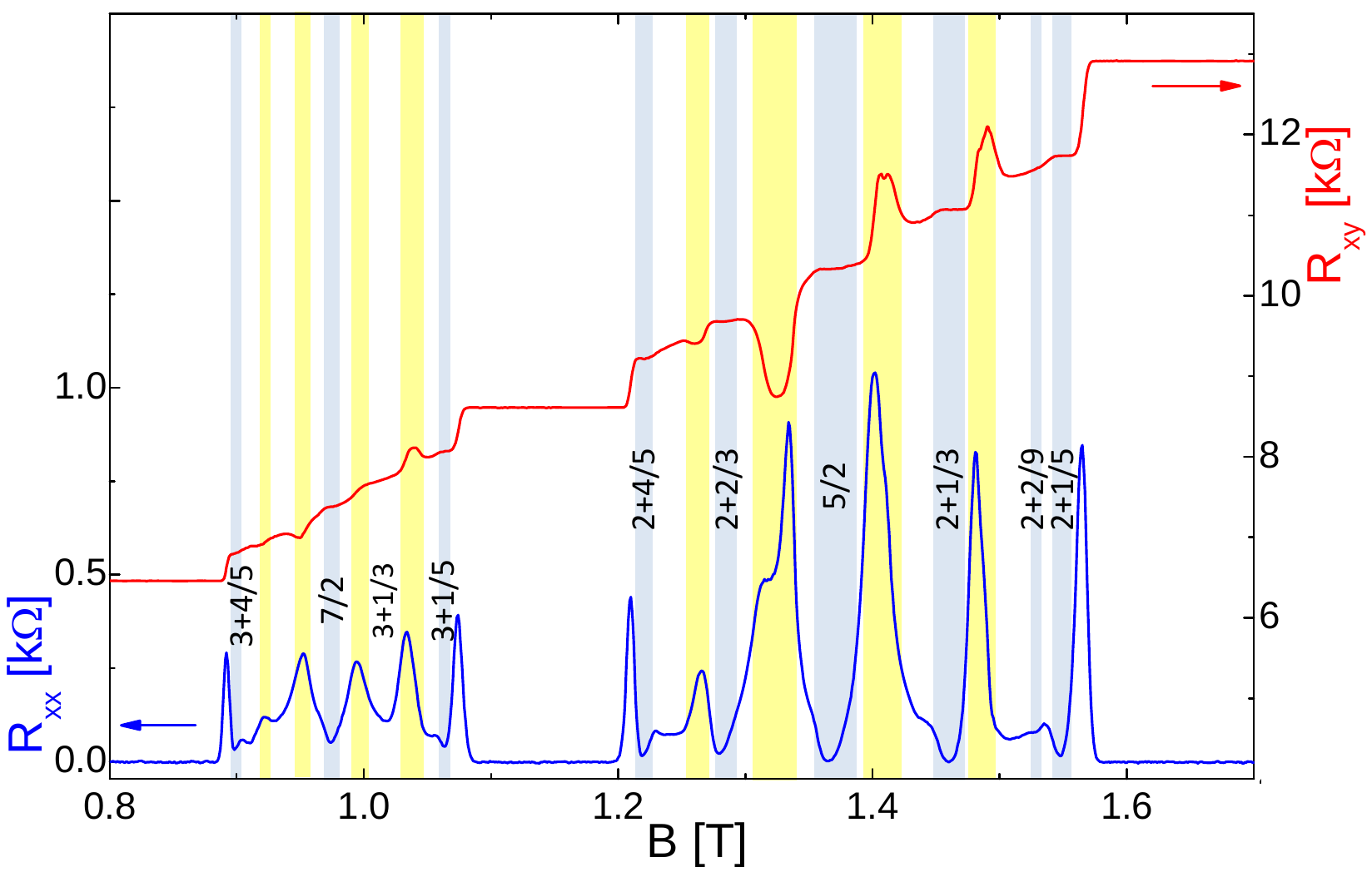}\label{FigC-8}}
}
\end{figure}

In order to gain insight into FQHSs at partial filling $1/3$,
one may plot the energy gaps of these FQHSs against
the Coulomb energy $E_c$. As showwn in Fig.\ref{FigC-7},
the energy gaps of the FQHSs at $\nu=2+1/3$, $2+2/3$, and $3+1/3$
plotted this way follow a linear trend.  
This suggests that these three FQHSs are similar in nature and, in the sample studied,
are not close to a possible spin transition\cite{pan83}. 

The relationship between gaps and the Coulomb energy shown in Fig.\ref{FigC-7}
may be relevant to the absence of  FQHSs in the third Landau level.
As shown in  Fig.\ref{FigC-7}, the energy gaps at the Coulomb energies calculated 
at $\nu=3+2/3$, $4+1/3$, and $4+2/3$ extrapolate to a negative values. 
Within the phenomenological model embodied in Eq.(\ref{eq1}),
a disorder broadening exceeding the intrinsic gap results
in a negative measured gap. Under such circumstances the formation of a FQHS is forbidden.
Based on such an argument, in 
the sample under investigation disorder effects are too strong
for the observation of the opening of an energy gap at $\nu=3+2/3$, $4+1/3$, and $4+2/3$. 
According to the Hartree-Fock theory\cite{fogler,moessner},
exact diagonalization\cite{haldane-bubble},
and density matrix renormalization group calculations\cite{yoshi-bubble},
the formation of the RIQHSs in the third Landau level is usually
attributed to their favorable cohesion energy as compared to
that of FQHSs at partial filling $1/3$. 
However, it is also possible that FQHS of the third Landau level 
are suppressed solely  because of the presence of disorder.

The most surprising feature of magnetotransport in the upper spin branch 
is the relative robustness of the FQHS at $\nu=3+1/5$ when compared to that of the FQHS at $\nu=3+1/3$\cite{ethan15}.
This result is consistent with earlier work in which a FQHS was seen at $\nu=3+1/5$, but not at $\nu=3+1/3$\cite{eisen02}.
The energy gaps were found to obey the $ \Delta_{3+1/3} < \Delta_{3+1/5}$ relation\cite{ethan15}. 
This relationship is very unusual since an opposite inequality $ \Delta_{2+1/3} > \Delta_{2+1/5}$ holds
in the lower spin branch of the second Landau level\cite{pan99,choi08,dean08,kumar10,nuebler10}
and $ \Delta_{1/3} > \Delta_{1/5}$ in the lowest Landau level\cite{willett1,mallett}. 
The latter two relationships are often ascribed
to the robustness of flux-two composite fermions when compared to the higher order flux-four
objects\cite{jainBook}.
To summarize, the expected relationship between the energy gaps of the 
FQHSs at partial filling factors $1/3$ and $1/5$ is reversed
in the upper spin branch of the second Landau level\cite{ethan15}. This anomalous gap
reversal indicates an unanticipated difference between the prominent odd denominator
FQHSs forming in the second Landau level\cite{ethan15}.

Since the upper spin branch forms at higher values of the Landau level mixing parameter
than the lower spin branch, the mixing effect is likely of importance. However, details of
the influence of Landau level mixing are not understood. To illustrate this, 
magnetoresistance is shown  in Fig.\ref{FigC-8} for a sample of a low electron density
$n=8.3 \times 10^{10}$~cm$^{-2}$. There are prominent
minima developed in the magnetoresistance and a strong Hall quantization at $\nu=3+1/5$ and $3+4/5$, 
but only a weak minimum and a lack of Hall quantization at $\nu=3+1/3$.
These results suggest that FQHSs at partial fillings $1/5$ stay
strong in the upper spin branch of the second Landau level
even at significantly large values of the Landau level mixing parameter.

The above described anomalous energy gaps of the prominent odd denominator FQHSs 
in the upper spin branch of the second Landau level
highlight the lack of understanding of these FQHSs and even elicit the provocative
possibility that some of these FQHSs may be of exotic origin
\cite{ethan15}.

\subsection{$\nu=4/11$ and $5/13$ fractional quantum Hall states}

One of the regions of interest for the observation of novel FQHSs in the
lowest Landau level is that between $1/3 < \nu <2/5$.
The observation of a local minimum in the magnetoresistance in this range at $\nu=4/11$ 
and of less pronounced features at filling factors such as $\nu=5/13$,
$6/17$, and $3/8$ was an early indicator of possible fractional quantum Hall ground states
\cite{WPan03}.
Initially the FQHS at $\nu=4/11$ was interpreted as a fractional quantum Hall effect of composite fermions\cite{WPan03}.
However, theoretical work suggested that, owing to the residual interactions between the composite fermions,
at $\nu=4/11$  the ground state is a FQHS with an unusual topological order\cite{Sitko,AWojs04,AWojs07}. 
This idea received strong support
from a composite fermion diagonalization study over an extended Hilbert space\cite{mukherjee14}.

\begin{figure}[t]
\centerline{
     {\includegraphics[width=2.5in]{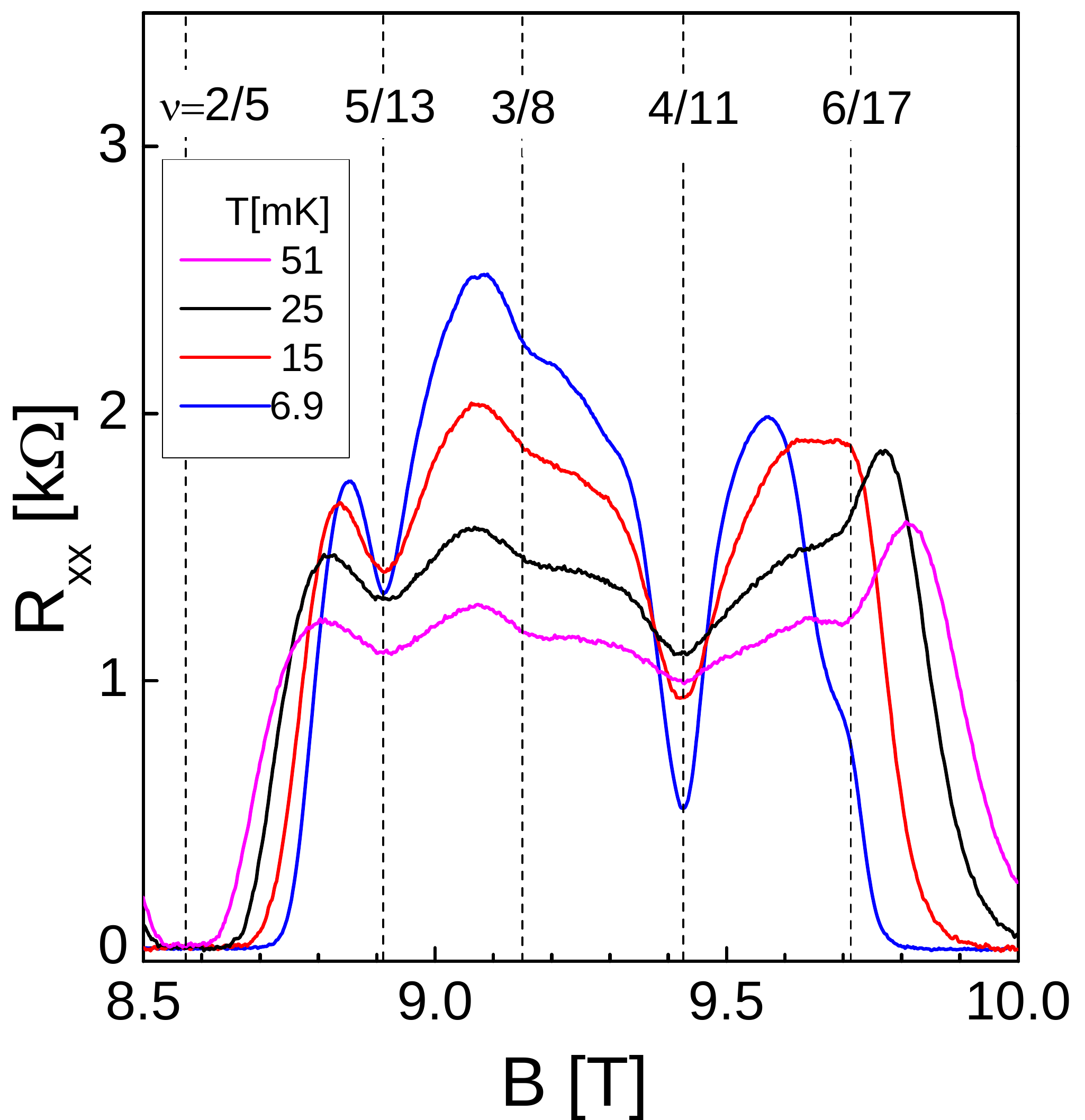}    }
}
\caption{Magnetotransport in the range of filling factors $1/3<\nu<2/5$ of the lowest Landau level.
   Developing FQHSs are seen at $\nu=4/11$ and $5/13$.
   Adapted from Ref.\refcite{nodar15}.} \label{FigC-9}
\end{figure}

While the theory results favoring a FQHS with a novel topological order at $\nu=4/11$ 
are compelling \cite{AWojs04,AWojs07,mukherjee14}, 
the lack of observation of an energy gap in the excitation spectrum, i.e.
of incompressibility, at this and other nearby filling factors raised the question
whether fractional quantum Hall ground states can be experimentally realized.
A necessary condition for the opening of a gap is a longitudinal magnetoresistance 
$R_{xx}$ that decreases with temperature.

The most recent experimental work in the $1/3 < \nu < 2/5$ range focused on searching for
an activated behavior. There are two reports that confirmed the
development of Hall quantization and the opening of an energy
gap at $\nu=4/11$, of magnitudes $7$~mK\cite{pan15} and $15$~mK\cite{nodar15}, respectively.
Magnetotransport data in the $1/3 < \nu < 2/5$ range from Ref.\refcite{nodar15}
are shown in Fig.\ref{FigC-9}.
Temperature dependence of magnetotransport at $\nu=5/13$ was also found to be
consistent with the development of an incipient gap\cite{nodar15}.
These measurements established fractional quantum Hall ground states at $\nu=4/11$ and $5/13$
and opened up the possibility for novel topological order in the $1/3 < \nu <2/5$ region.

Despite considerable progress at $\nu=4/11$ and $5/13$, 
transport data at other filling factors of interest, such as $\nu=3/8$ and $6/17$, 
did not conclusively establish the opening of an energy gap\cite{pan15,nodar15}.
Magnetoresistance at $\nu=3/8$ in both experiments
 was found to increase with a lowering temperature, precluding therefore 
activation. Furthermore, the filling factor $\nu=6/17$ is extremely close to the quantized plateau associated
with the $\nu=1/3$ FQHS.
A recent numerical study of the FQHS at $\nu=6/17$ found it is described by the model
of non-interacting composite fermions\cite{ajit16}. 
For now, the behavior of the magnetoresistance at $\nu=3/8$ and $6/17$ 
cannot be conclusively associated with the opening of an energy gap  
despite the existence of a depression in $R_{xx}$ at these filling factors.

\section{Even denominator fractional quantum Hall states}
\label{cs-sec4}

The FQHS at $\nu=5/2$ is the
most notable FQHS beyond the model of non-interacting composite fermions. 
This FQHS was discovered\cite{willett1} and its full quantization was 
reported in Refs.\refcite{jim90,pan99}. Related FQHSs develop at
other filling factors of even denominator in the GaAs/AlGaAs system
at $\nu = 7/2$\cite{eisen02} and $\nu = 2 + 3/8$\cite{xia04}.
Even denominator FQHSs have also been seen in ZnO/MgZnO\cite{falson}
and in bilayer\cite{gr0,gr1,gr2,gr3} and, most recently, in monolayer\cite{gr4,gr5} graphene. 
However, it is not yet established whether or not these latter FQHSs and the $\nu=5/2$ FQHS in GaAs/AlGaAs
share the same origin.

The instability of the Fermi sea of composite fermions and the
existence of an energy gap at this filling factor is naturally explained by
a Cooper-like pairing of the composite fermions\cite{mr,pair1,pair2,pair3,pair4,pair5,pair6}. 
Because of constraints
on the spin degree of freedom, such a pairing must necessarily be $p$-wave pairing\cite{pair2}.
The Pfaffian is the earliest description of the FQHS at $\nu=5/2$ which is consistent with such a pairing\cite{mr}.
The Pfaffian is of considerable interest since at least some of its quasiparticles are predicted to 
obey exotic non-Abelian braiding statistics.
However, at this filling factor there are also other topologically distinct candidates.
such as  the anti-Pfaffian\cite{anti1,anti2}, the (3,3,1) Abelian state\cite{halperin331},
a variational wave function based on an anti-symmetrized
bilayer state\cite{anti3}, the particle-hole symmetric Pfaffian\cite{dirac1,dirac2},
a stripe-like alternation of the Pfaffian and anti-Pfaffian\cite{anti4},
and other exotic states\cite{wen90,wen91}. 

An intense effort is focused on unraveling the properties of the even denominator FQHSs. 
Some aspects of this effort were reviewed in Refs.\cite{willettRev,duRev}.
Edge-to-edge tunneling \cite{tunn1,tunn2,tunn3,tunn4}, quasiparticle interferometry\cite{interfero},
upstream neutral modes\cite{neutral}, edge heat conduction measurements\cite{heat} 
probed the structure of the edge states at $\nu=5/2$. However, results from these measurements
do not yet offer a consensus on the origin of the $\nu=5/2$ FQHS.
Since the physics of the edge may be considerably more complicated than that
of the bulk, bulk probes such as transport and heat capacity measurements\cite{gerv}
remain important in the study of the FQHS at $\nu=5/2$.
In this section we discuss results on the $\nu=5/2$ FQHS that were learnt
from recent transport data. Several experiments suggest that away from the optimal density
phase transitions are allowed in the FQHS at $\nu=5/2$. We will also discuss
the energy gap and the disorder broadening of the FQHS at this filling factor
in the highest quality samples. The last subsection contains results of a systematic
study of the behavior of the $\nu=5/2$ FQHS in the presence of short-range alloy disorder.

\subsection{$\nu=5/2$  fractional quantum Hall state at low electron densities}

The regime of low densities for the $\nu=5/2$ FQHS emerged as an area of interest
because of the possibility of phase transitions occurring here. In the following we discuss
a spin transition\cite{pan14} and a topological phase transition at $\nu=5/2$\cite{nodar17}. 
In addition, a transition from the $\nu=5/2$ FQHS towards 
the quantum Hall nematic will be discussed in \sref{cs-sec6}\cite{kate1,kate2,kate3}.
 
A possible phase transition in the $\nu=5/2$ FQHS is a spin transition, from a fully spin polarized 
to a partially polarized state.
According to experimental work performed at large magnetic fields and hence large densities, 
the $\nu=5/2$ FQHS is fully spin polarized\cite{tiemann,ursula}.
However, these experiments do not rule out a partially spin polarized FQHS
at low electron densities.
Recent experimental work on a series of samples, including samples of density as low as
$n=4.1 \times 10^{10}$~cm$^{-2}$, found a region of densities at which the
energy gap nearly closes\cite{pan14}. These results have been
interpreted as being consistent with a spin transition in the $\nu=5/2$ FQHS\cite{pan14}.
While these results\cite{pan14} are suggestive of a phase transition, a spin transition is only one 
of the possibilities. Indeed, while changing the density other parameters of the system may also change.
Examples of such parameters are the Landau level mixing parameter,
defined as the ratio of Coulomb and cyclotron energies,  and the width of the quantum well.
Therefore orbitally driven phase transitions may also occur when the density is changed.

\begin{figure}[t]
\centerline{\includegraphics[width=1.97in]{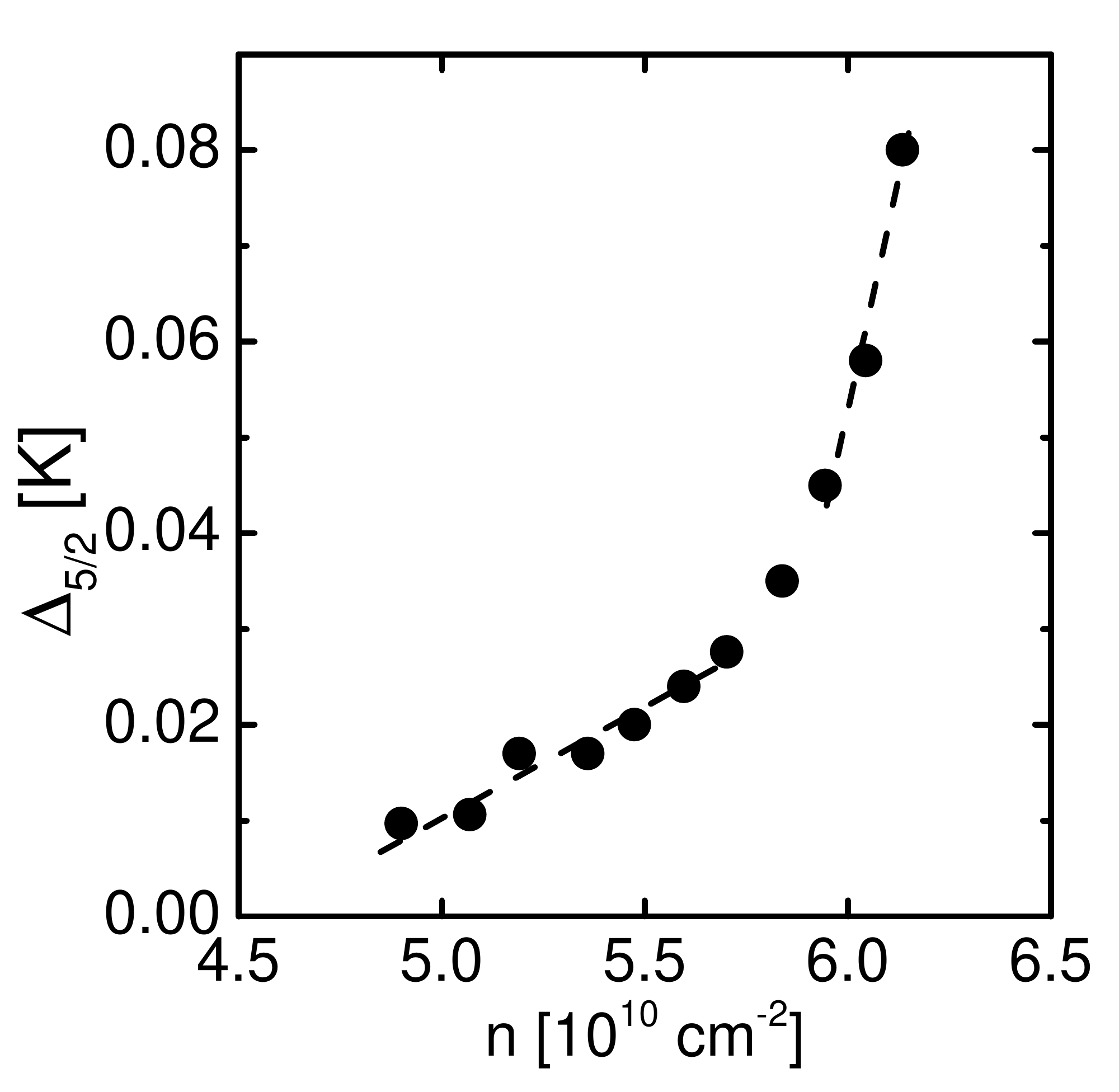}}
\caption{Energy gap of the $\nu=5/2$ FQHS versus density in a high quality low density sample.
      The anomalous trend of the energy gap at $\nu=5/2$ versus density may be interpreted as being due to 
      a topological phase transition. Adapted from Ref.\refcite{nodar17}.}
\label{FigC-10}
\end{figure}

As discussed earlier, theory finds that for the $\nu=5/2$ FQHS
several topologically different candidate states are allowed \cite{mr,anti1,anti2,halperin331,anti3,dirac1,dirac2,anti4,wen90,wen91}. 
If more than one of these states can be stabilized, intriguing topological phase transitions may occur
at $\nu=5/2$ between pairs of such distinct FQHSs.
Such phase transitions may be driven by a parameter of the 2DEG, such as
the  Landau level mixing parameter $\kappa$. As an example, it was
argued that a direct topological phase transition between the
Pfaffian and the anti-Pfaffian may occur\cite{anti1,anti2}. According to numerical work,
the Pfaffian and anti-Pfaffian ground states may compete as the Landau level mixing 
 is tuned\cite{sim1,sim2,sim4,sim5,sim6,sim7,sim8,sim9,papic09,chakra17}. 
However, due to difficulties stemming from the non-perturbative nature of the calculations
and due to limited computational resources, details of a possible transition between
the Pfaffian and the anti-Pfaffian at large Landau level mixing could not be firmly
established\cite{sim9}. 
Nonetheless, the regime of low densities or large Landau level mixing has
emerged as a region of interest for a possible topological phase transition
in the $\nu = 5/2$ FQHS.

In a recent experiment on a density-tuned 2DEG\cite{nodar17},
an anomalously sharp change in the density dependence of
the energy gap of the $\nu = 5/2$ FQHS was reported 
in the vicinity of $\kappa = 2.6$ or $n = 5.8 \times 10^{10}$~cm$^{-2}$.
This behavior is shown in Fig.\ref{FigC-10}.
The origins of the observed anomalous dependence of the energy gap are not
known; one possibility is a topological phase transition in the $\nu=5/2$ FQHS\cite{nodar17}.
We note that the energy gap at the apparent transition point in this experiment
did not close. While the closure of the gap is generally believed to be
a necessary condition for a topological phase transition,
one may also envision topological transitions in which the gap does not fully close,
for example in a first order phase transition.
A topological phase transition in the $\nu=5/2$ FQHS may also be induced in confined geometries
by changing the confinement potential\cite{tunn4,bo}.

Another interesting possibility for a phase transition at $\nu=5/2$ from a FQHS towards the
quantum Hall nematic
was recently revealed by measurements at high hydrostatic pressures\cite{kate1,kate2}. 
Further details of this phase transition can be found in \sref{cs-sec6}.

\subsection{Energy gap at $\nu=5/2$ in pristine samples}

Unraveling the effects of the disorder is an important endeavor in contemporary condensed matter physics.
Disorder is well understood in the single particle regime for example in connection with
Anderson localization and the universal plateau-to-plateau transition
in the integer quantum Hall effect\cite{wanli,wanli2}.  
In contrast, understanding disorder in correlated electron systems, such
as the 2DEG in the fractional quantum Hall regime, continues to pose serious challenges. 

In the following, we will refer to samples of the highest possible mobility and, therefore, the least
amount of disorder as pristine samples. The highest mobility under given growth conditions
is a function of the electron density\cite{loren1}.

Efforts in understanding the energy gap of the $\nu=5/2$ FQHS examined its relationship to
electron mobility $\mu$ and
quantum lifetime $\tau_q$. In the most general case, the energy gap at $\nu=5/2$
in pristine samples did not correlate with the mobility\cite{nodar11,pan08,dean08,nuebler10,gaas1,masu,gamez,nakamura}. 
Similarly, the energy gap at $\nu=5/2$ did not scale with the quantum lifetime\cite{nodar11,qian17}. 
In fact in gated samples it was found that the quantum lifetime
is approximately constant over the density range at which the energy gap at $\nu=5/2$ 
decreased from its largest value to zero\cite{nuebler10,qian17}.
These results are perhaps not surprising since, in contrast to the energy gap at $\nu=5/2$,
both the mobility and the quantum lifetime are parameters measured
near zero magnetic field, in a regime that may be understood within a single electron description.

The phenomenological model described earlier in \sref{cs-sec3p1}
offers a framework for analyzing the gap of the $\nu=5/2$ FQHS. The use of this model
for gaps measured in the second Landau level, however, poses challenges
as these gaps were suspected to be smaller than the disorder broadening.
In contrast, $\Gamma$ at $\nu=1/3$ is neglijible as compared to $\Delta^{int}$ and
the intrinsic gap may be estimated from the measured value of the gap, $\Delta^{int} \simeq \Delta$ 
in samples of high density \cite{dis1,dis2}. Therefore in order to estimate
$\Delta^{int}$ at $\nu=5/2$, one needs to measure 
independently two quantities: both $\Delta$ and $\Gamma$. 
Hence a quantitative knowledge of the the disorder broadening plays a significant role in 
understanding the energy gaps of the exotic FQHSs.
However, $\Gamma$ is not directly accessible from gap measurements at $\nu=5/2$.

\begin{figure}[t]
\centerline{\includegraphics[width=0.96\columnwidth]{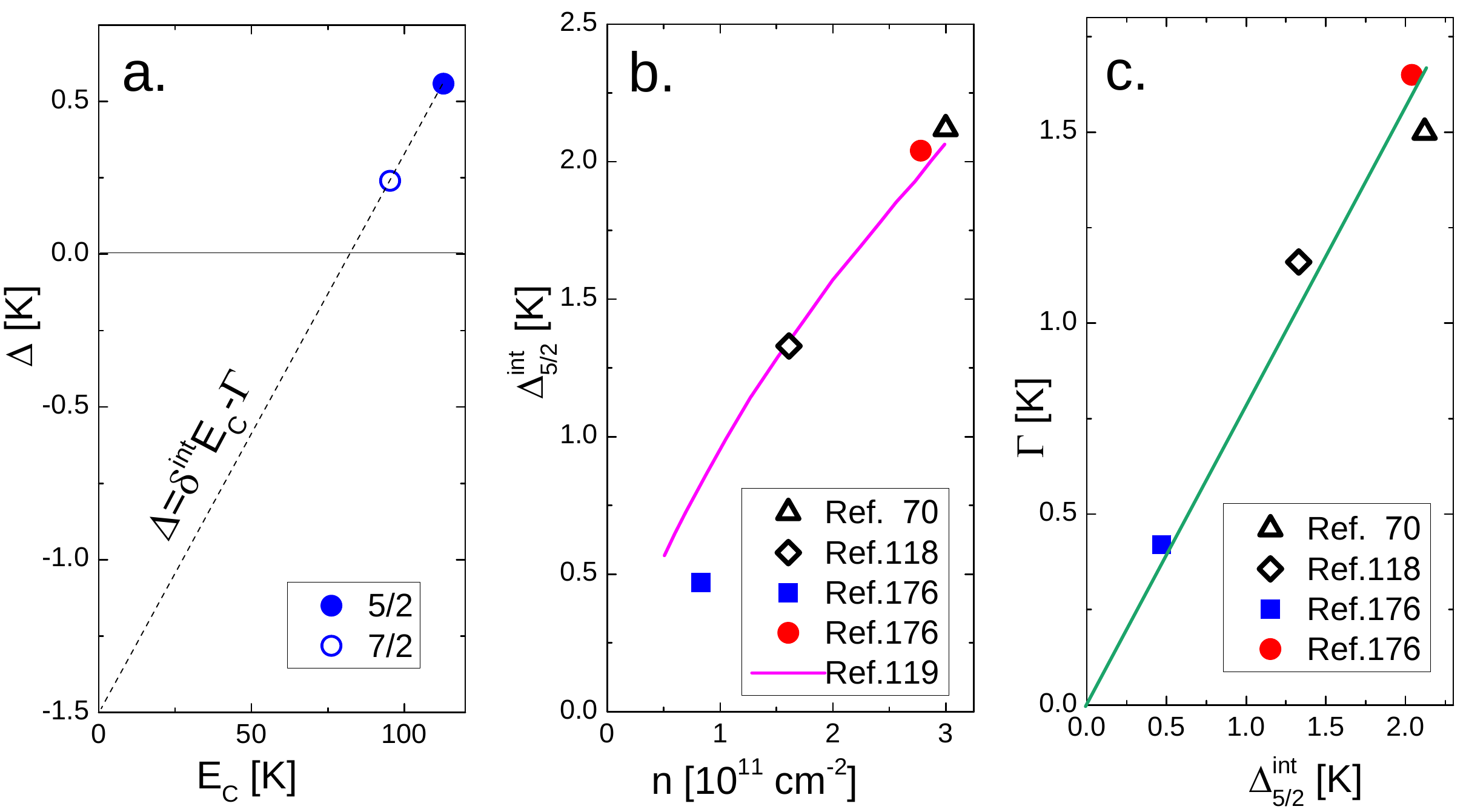}}
\caption{Panel a: A fit to the measured gaps at even denominators used to extract the dimensionless intrinsic gap 
$\delta^{int}$ and the disorder broadening $\Gamma$. Data is from Ref.\refcite{nodar11}.
Panel b: A comparison of intrinsic gaps at $\nu=5/2$
obtained from measurements (data points) and numerics (line). Data taken from Ref.\refcite{nodar11}.
Panel c: The dependence of $\Gamma$ on $\Delta^{int}_{5/2}$.}
\label{FigC-11}
\end{figure}

To resolve this impasse, Morf and d'Ambrumenil\cite{ambru} proposed an analysis based on
the measurement of two independent quantities: the energy gaps of FQHSs at both $\nu=5/2$ and $\nu=7/2$
in a sample of given electron density. 
In addition, it was assumed that these two FQHSs have the same dimensionless
intrinsic gap $\delta^{int} = \Delta^{int} / E_c$, where  $E_c$ is
the Coulomb energy.
The dimensionless intrinsic gap $\delta^{int}$ and disorder
broadening $\Gamma$ may be obtained by plotting the measured gaps at $\nu=5/2$
and $7/2$ against the Coulomb energy and by extracting the slope and the intercept with the vertical scale
of the line passing through the two points. Following this recipe\cite{ambru}, an analysis 
of the gaps in a sample of electron density
$n=3.0 \times 10^{11}$~cm$^{-2}$ from Ref.\refcite{eisen02} yielded values
$\delta^{int}=0.014$ and $\Gamma=1.24$~K. A similar fit, shown in Fig.\ref{FigC-11}a, 
on the sample from Ref.\refcite{kumar10} yielded $\delta^{int}=0.019$ and $\Gamma=1.5$~K\cite{nodar11}.

The analysis of the gaps proposed by Morf and d'Ambrumenil\cite{ambru}
was later extended for samples of low densities \cite{nodar11}, down to $n=8.3  \times 10^{10}$~cm$^{-2}$.
The intrinsic gaps as plotted against the density, shown in Fig.\ref{FigC-11}b, 
were in reasonable agreement with a numerical simulation 
that accounted for Landau level mixing within the random phase approximation\cite{nuebler10}.
However, such an agreement can only be considered crude at best since the
assumption of equal $\delta^{int}$ for the $\nu=5/2$ and $\nu=7/2$ FQHSs is only
approximate in measurements due to the slightly different Landau level mixing parameters
at these filling factors.

\begin{figure}[t]
\centerline{
  \minifigure[Dimensionless intrinsic gaps at $\nu = 5/2$ as function of the Landau level
     mixing parameter $\kappa$. Dotted line is a fit through the data. Adapted from Ref.\refcite{nodar11}.]
     {\includegraphics[width=2.55in]{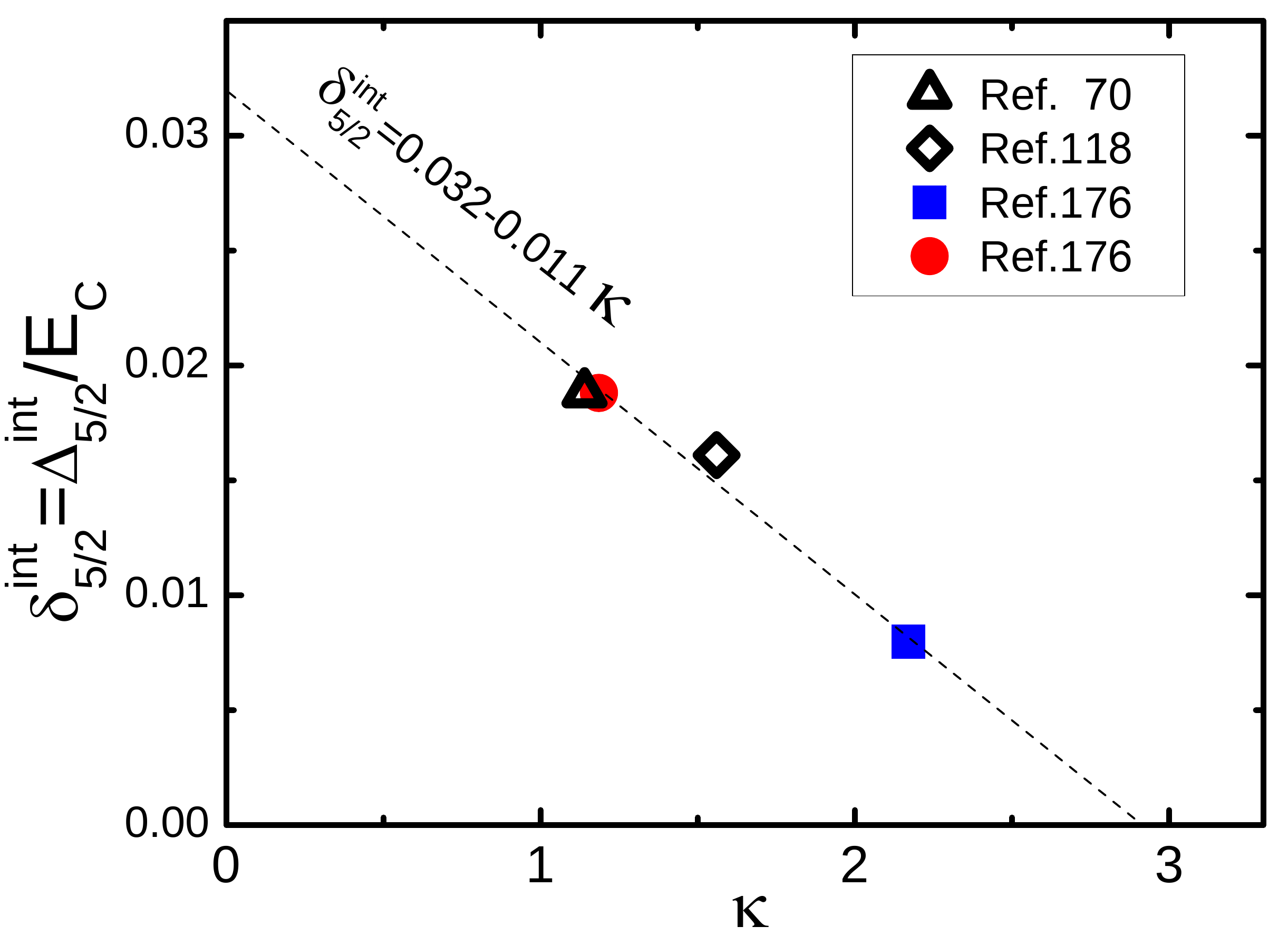}\label{FigC-12}}
  \hspace*{0pt}
  \minifigure[Magnetotransport in the $2 < \nu < 3$ range of a $p$-type sample. 
     A FQHS is seen at $\nu=2+2/3$. Adapted from Ref.\refcite{ptype}.]
     {\includegraphics[width=2.05in]{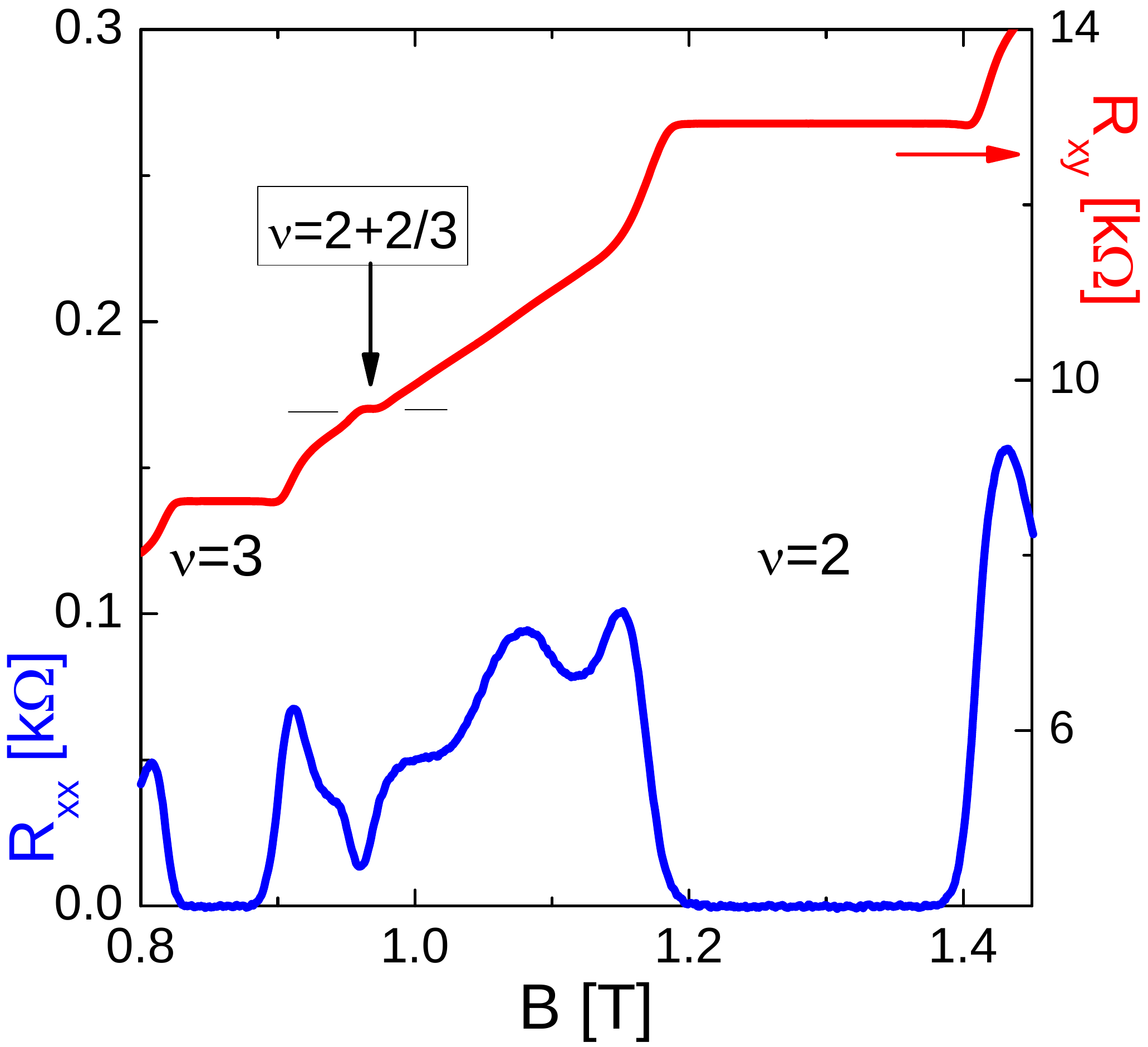}\label{FigC-13}}
}
\end{figure}

It is instructive to plot the calculated $\delta^{int}$ against the Landau level mixing parameter $\kappa$.
The decreasing trend obtained is shown in Fig.\ref{FigC-12}.
It was found that a linear fit to the data extrapolates to $\delta^{int}=0.032$ in the limit of $\kappa=0$.
This extrapolated value is in good agreement with the numerically obtained values at zero
Landau level mixing: $0.02-0.04$ from an exact diagonalization on finite size systems\cite{morf},
$0.030$ from a DMRG calculation\cite{feiguin}
and $0.031$ from an exact diagonalization study adjusted for the finite width of the quantum well\cite{nuebler10}.
The latter two figures are from extrapolations to the thermodynamic limit.
Since simulations with no Landau level mixing are significantly easier than the ones that include it\cite{sim9}, 
the agreement of $\delta^{int}$ from these simulations and $\delta^{int}$
extrapolated to $\kappa=0$ is an important confidence test of the analysis above.

Another extrapolation of interest is to large $\kappa$. The functional form for this
extrapolation is not known; in Ref.\refcite{nodar11} a linear extrapolation was used. 
The extrapolated intrinsic gap at $\nu=5/2$
was found to close near $\kappa \simeq 3$\cite{nodar11}. This result is consistent with the
totality of observations of electron gases at low densities in the GaAs/AlGaAs system\cite{nodar11,pan14}. 
This result is also consistent with observations in $p$-type carriers, i.e. two dimensional hole gases in GaAs/AlGaAs
in which values of $\kappa$ are significantly larger than 3 and in which an even denominator FQHS
was not observed. As an example, in Fig.\ref{FigC-13} we show the $2<\nu<3$ range
for a hole gas with $\kappa = 14$, 
in which an odd denominator FQHS develops, but an even denominator FQHS is not
present\cite{ptype}.
We note that it is widely appreciated that, besides Landau level mixing, energy gaps
also depend on the width of the quantum well in its dimensionless form $w/l_B$.
For the samples on which Fig.\ref{FigC-12} is based on, these dimensionless widths
were fortuitously nearly constant\cite{nodar11}.

Progress in MBE technology has resulted in a dramatic improvement of the quality of 2DEG confined
not only to GaAs/AlGaAs\cite{loren1}, but also
to ZnO/MgZnO structures\cite{semi5}. In such 2DEGs of increased mobility
there is a recent report\cite{falson} of even denominator FQHSs. 
Measured parameters of the 2DEG in this experiment show that these even denominator FQHSs
occur near $\kappa \simeq 15$\cite{falson}. 
According to the extrapolation to large $\kappa$ shown in  Fig.\ref{FigC-12},
at such large values of $\kappa$ a FQHS at $\nu=5/2$ cannot be stabilized in GaAs/AlGaAs samples.
This result thus raises the interesting possibility
that the even denominator FQHSs in GaAs/AlGaAs and in ZnO/MnZnO may have a different origin.

Finally, it is interesting to note that, as seen in Fig.\ref{FigC-11}c,
the disorder broadening $\Gamma$ in these samples\cite{nodar11}, extracted
by the technique of Morf and d'Ambrumenil, is found to be linear with the intrinsic gap $\Delta^{int}$.
The slope of a linear fit to data in Fig.\ref{FigC-11}c passing through the origin is $0.78$.

\subsection{Energy gap at $\nu=5/2$ in samples with alloy disorder}

Progress in MBE growth of GaAs/AlGaAs structures has produced sufficiently high quality
2DEGs to observe quantum Hall phenomena
even when there was disorder intentionally added to the 2DEG. Adding disorder
during the MBE growth process allowed for a high degree of its control. 
So far only alloy disorder was systematically studied\cite{wanli0,alloy1}. 
The particular type of alloy disorder investigated consisted of
minute amounts of neutral Al atoms added to the GaAs region supporting the 2DEG.
Experiments on such alloy samples had a strong impact on the understanding
of the plateau-to-plateau transition in the integer quantum Hall regime\cite{wanli,wanli2},
mapping the electronic wavefunction along a direction perpendicular to the plane of the 2DEG\cite{wegscheider},
and contributed to understanding of pinning of Wigner crystals\cite{moonengel}.

\begin{figure}[b]
\centerline{
  \minifigure[The dependence of the scattering rate $1/\tau$ on alloy disorder
     in a series of 30~nm quantum well based alloy samples. 
     Here $x$ is the Al molar concentration. Adapted from Ref.\refcite{alloy1}.]
     {\includegraphics[width=2.3in]{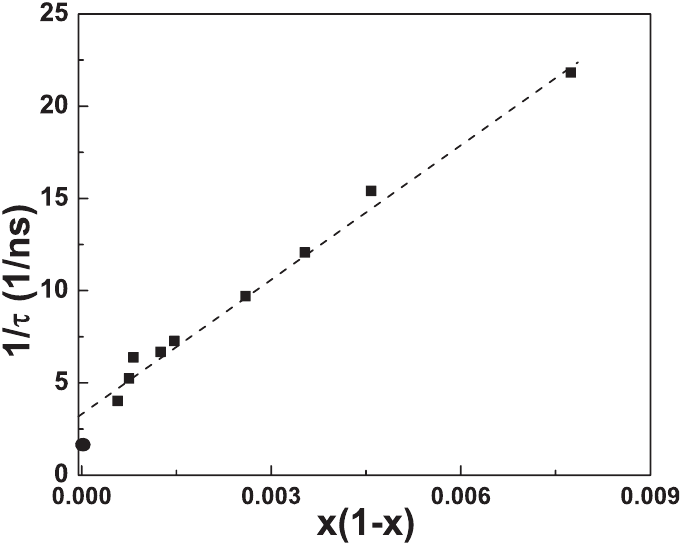}\label{FigC-14}}
  \hspace*{0pt}
  \minifigure[The dependence of the energy gap of the $\nu=5/2$ FQHS on inverse mobility
     in a set of pristine and a series of alloy samples near the density $3 \times 10^{11}$~cm$^{-2}$.
     Adapted from Ref.\refcite{alloy2}.]
     {\includegraphics[width=2.3in]{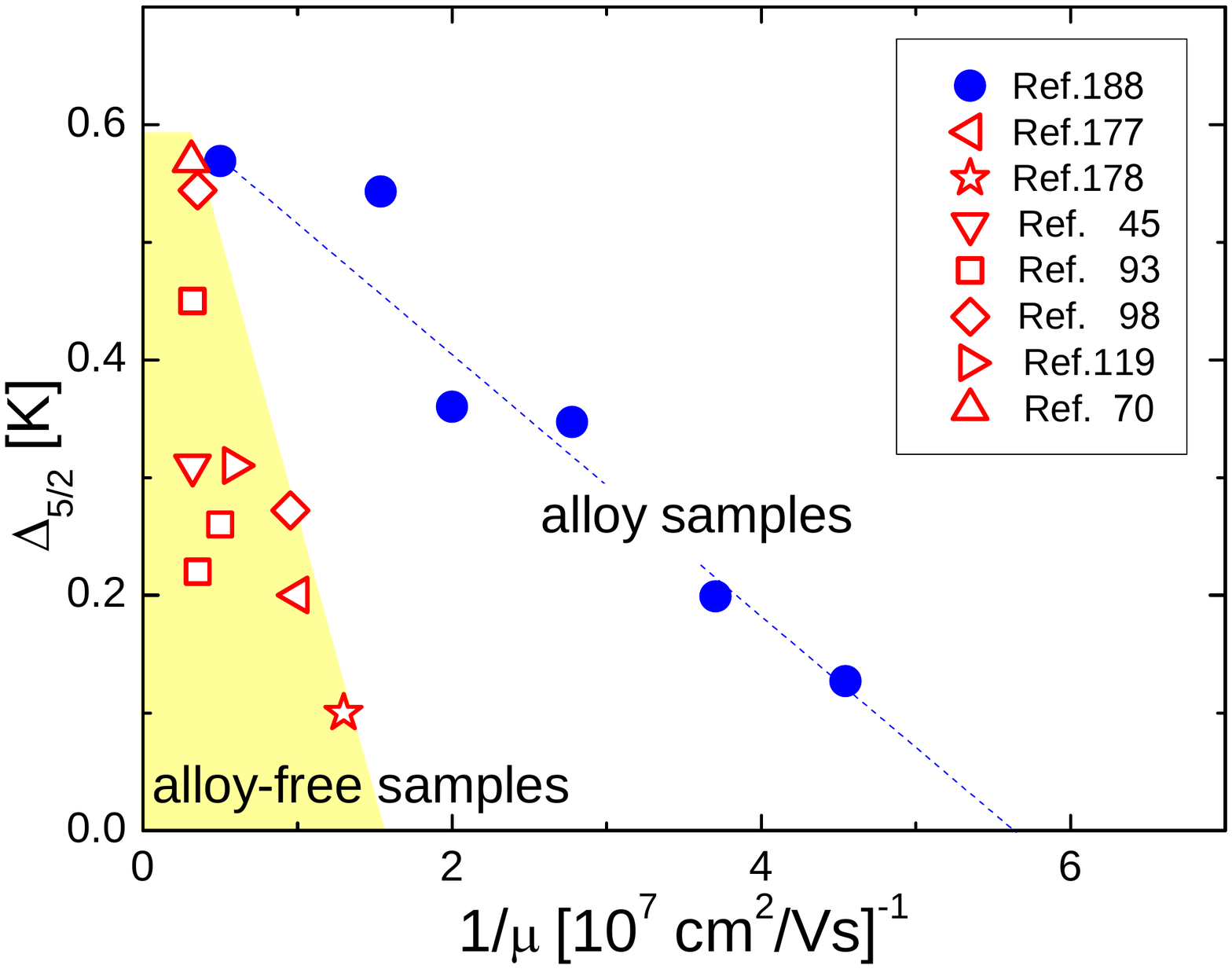}\label{FigC-15}}
}
\end{figure}

Another successful experiment was performed on alloy samples that enabled
the study of the FQHS at $\nu=5/2$ in the presence of alloy
disorder\cite{alloy1,alloy2}. 
These samples were based on modern sample structures assuring the high density, 
near $n \simeq 3 \times 10^{11}$~cm$^{-2}$, and high quality necessary
for robust second Landau level fractional quantum Hall states. Specifically, samples were
modulation-doped Al$_{0.24}$Ga$_{0.76}$As/Al$_x$Ga$_{1-x}$As/Al$_{0.24}$Ga$_{0.76}$As 
quantum wells, where the molar Al fraction
$x$ in the quantum well was significantly less than that in the barriers\cite{alloy1}.
The dependence of the scattering rate on the amount of Al introduced
to the quantum well can be seen in Fig.\ref{FigC-14}. 
The linear dependence and slope of the scattering rate on the amount of Al
characterizes the alloy potential and was
was found to be consistent with earlier results\cite{wanli0}. Moreover, because of the small scattering rate
in the pristine sample with no added impurities to the quantum well, alloy scattering exceeded
the residual scattering rate
in all these samples, with the exception of the one with the lowest non-zero $x$\cite{alloy1}.

As shown in Fig.\ref{FigC-15}, an increasing amount of alloy disorder decreases the mobility and it
also suppresses the energy gap of the $\nu=5/2$ FQHS\cite{alloy2}. A similar behavior
was found in numerical work on the $\nu=1/3$ FQHS, in which the strength of  a Gaussian correlated 
white noise was increased\cite{disord1,disord2,disord3,disord4,alloy4}. 
A peculiarly interesting feature of the data is revealed when the alloy samples are compared
to high quality pristine, i.e. alloy-free samples. Because the mobility is not independent
of the sample density, in Fig.\ref{FigC-15} only pristine samples from the literature are shown
that have their density in the
$2.65 \times 10^{11} \leq n \leq 3.2 \times 10^{11}$~ cm$^{-2}$ range, i.e. 
close to that of the alloy samples. Energy gaps for the pristine samples are clustered in the area
shaded in yellow in Fig.\ref{FigC-15}, but shown no correlation with $1/\mu$. In contrast,
the energy gap in the series of alloy samples shows
a linear functional dependence on the inverse mobility $1/\mu$. 
This suggests that when in a series of similar samples one particular type of disorder dominates,
such as in the series of alloy samples, the energy gap and and the mobility appear to be correlated\cite{alloy2}.
However, in samples which have different types of  dominating disorder, a correlation between
the energy gap and and the mobility is not expected\cite{alloy2}.

An interesting feature of the data shown in  Fig.\ref{FigC-15} is that 
a strong $\nu=5/2$ FQHS with $\Delta_{5/2}=127$~mK develops
in the alloy sample with $\mu=2.2 \times 10^{6}$~cm$^2$/Vs. 
This is surprising, since at such a low mobility a $\nu=5/2$ FQHS has never been observed
in pristine, alloy-free samples. Indeed, for pristine, alloy-free samples with
density near $n=3 \times 10^{11}$~cm$^{-2}$, the threshold mobility value for
the observation of a $\nu=5/2$ FQHS is $\mu_c \simeq 7 \times 10^{6}$~cm$^2$/Vs.
The mobility threshold for a fully quantized $\nu=5/2$ FQHS is therefore significantly lowered in the presence of alloy disorder
and, therefore, the $\nu=5/2$ FQHS is robust to the presence of alloy disorder.\cite{alloy2} 
Alloy disorder does not appear to be as detrimental to the development of the 
$\nu=5/2$ FQHS as the residual disorder unintentionally added during sample growth\cite{alloy2}. 
The gap $\Delta_{5/2}$ in the series of alloy samples studied
closes at an extrapolated threshold of $\mu_c^{alloy} \simeq 1.8 \times 10^{6}$~cm$^2$/Vs.

Al is a neutral impurity and it perturbs the GaAs crystal on a subnanometer
length scale. Alloy disorder is thus a type of disorder that generates a short-range scattering potential.
Other types of disorder, due to either short- or long-range scattering potentials,
have not yet been systematically studied.

\section{Studies of reentrant integer quantum Hall states}
\label{cs-sec5}

RIQHSs are collectively localized ground states associated with electronic bubble phases\cite{fogler,moessner,fradkin,lilly99,du99}.
They were discovered in the third and higher Landau levels\cite{lilly99,du99}
and later also seen in the second Landau level\cite{eisen02}.
There are also two reports of RIQHSs in the lowest Landau level, however
the nature of these RIQHSs may be different from those forming in higher Landau levels\cite{lll-1,lll-2}.
In this section we discuss recent results on RIQHSs in the second Landau level. Topics include 
the magnetoresistive fingerprints of the RIQHSs in high mobility samples, 
a discussion of the precursors of the RIQHSs, and a summary of new RIQHSs.

Transport signatures of RIQHSs are $R_{xx}=0$
and $R_{xy}=h/i e^2$, where $i=$~integer\cite{lilly99,du99}.
However, in contrast to integer quantum Hall states, RIQHSs are centered at a filling factor
other than an integer. 
A characteristic of the RIQHSs is that they are separated from the nearby integer quantum 
Hall plateaus by distinct resistive signatures\cite{lilly99,du99}.
Regions shaded in yellow in Fig.\ref{FigC-4} and Fig.\ref{FigC-16}
mark several RIQHSs forming in the second Landau level. 
The above transport signatures of RIQHSs indicate a pinned insulator and the
reentrant property was argued to distinguish RIQHSs from an Anderson insulator\cite{cooper,eisen02}.

Following the discovery of RIQHSs, efforts on these states were focused 
on establishing their fundamental signatures in magnetotransport, on
their filling factors of formation\cite{eisen02,deng12,deng12b,shingla}, 
on gaining an understanding of their energy scale\cite{deng12,deng12b},
on  their microwave pinning modes\cite{lewis-1,lewis-2,lewis-3}, on thermopower\cite{chick13},
on their electrical breakdown\cite{cooper,break-1,break-2,break-3,break-4,break-5}, and most recently,
on their attenuation of surface acoustic waves\cite{dietsche,smet-saw}.

RIQHSs develop at temperatures higher than 100~mK in the third Landau level and
at tens of mK in the second Landau level. Therefore the use of 
a He-3 immersion cell to generate ultra-low temperatures may seem 
unnecessary in their study. Nonetheless,
immersion cells have contributed to the RIQHSs in an unexpected way: 
the large heat capacity of the He-3 liquid provided enhanced temperature stability. 
Since the magnetoresistance of RIQHS near their onset 
changes extremely rapidly with temperature,
the high temperature stability afforded by the immersion cell setup allowed for
a careful mapping of these temperature sensitive features\cite{deng12,deng12b,shingla}.

\subsection{Resistive fingerprints of reentrant integer quantum Hall states}

The vanishing longitudinal magnetoresistance and Hall quantization to an integer value
were transport features of RIQHSs clearly identified at their discovery\cite{lilly99,du99,eisen02}.
It later became apparent that samples with improved quality exhibit additional
transport features associated with RIQHSs.
One such feature is the flanking of the longitudinal magnetoresistance
by sharp peaks at both ends of the magnetic field. In other words, the region
of vanishing magnetoresistance of a RIQHS is delimited by two sharp peaks.
As seen in Fig.\ref{FigC-4}, 
these two sharp resistance peaks in the flanks of RIQHSs
are present in both spin branches of the second Landau level\cite{deng12}.
Similar sharp peaks are also delimiting the RIQHSs in the third Landau level (not shown)\cite{deng12b}.

\begin{figure}[t]
\centerline{
  \minifigure[Waterfall plot of magnetoresistance in the lower spin branch of the second Landau level,
     as measured at different temperatures. Traces are labeled by temperatures in mK. Adapted from Ref.\refcite{shingla}.]
     {\includegraphics[width=2.1in]{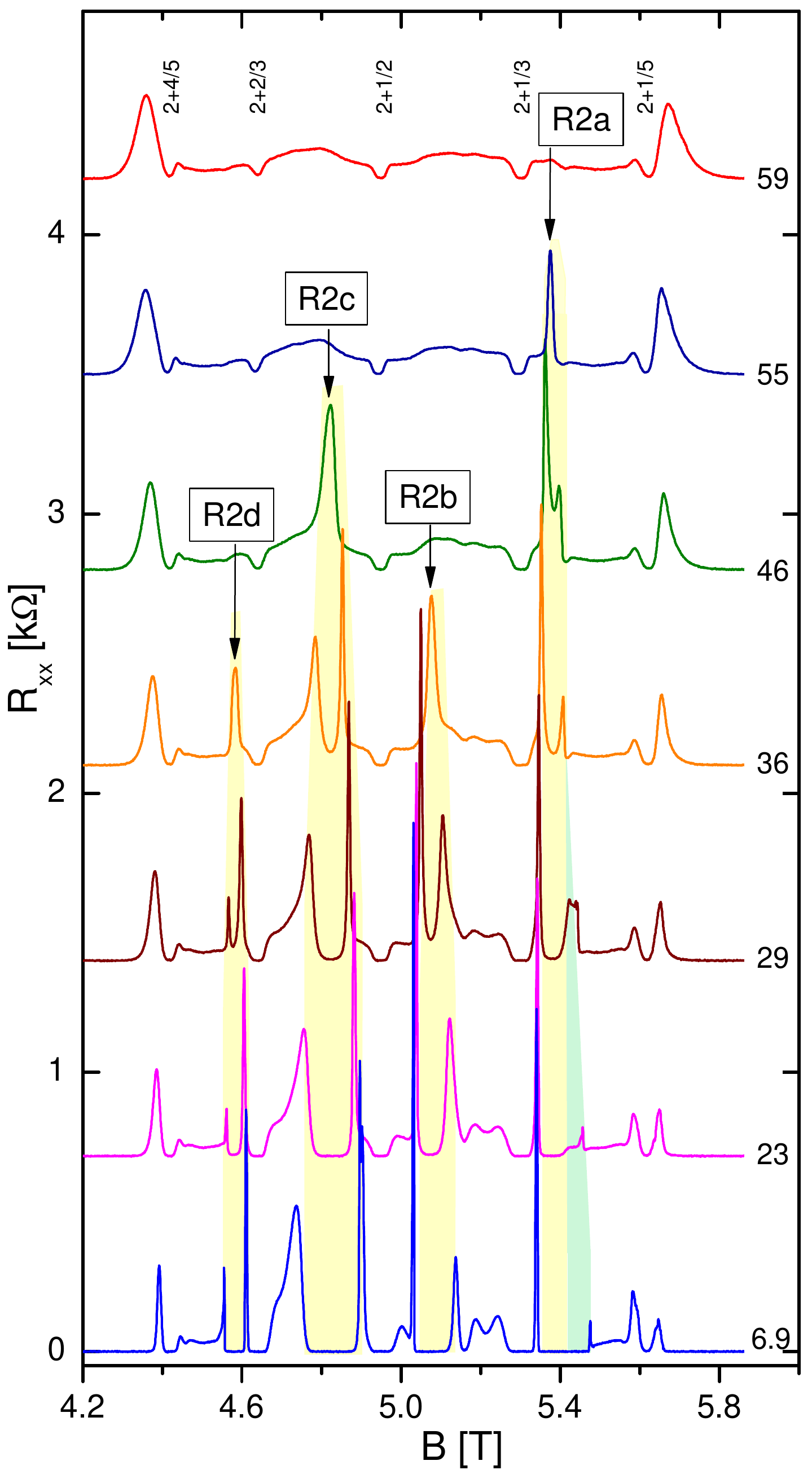}\label{FigC-16}}
  \hspace*{0pt}
  \minifigure[Temperature dependence of $R_{xx}$ and $R_{xy}$ measured at the central filling factor
     of the RIQHSs labeled $R2a$, $R2b$, $R2c$, and $R2d$. Sharp peaks in $R_{xx}$ mark the onset temperatures
     of the RIQHSs. Adapted from Ref.\refcite{deng12}.]
     {\includegraphics[width=2.42in]{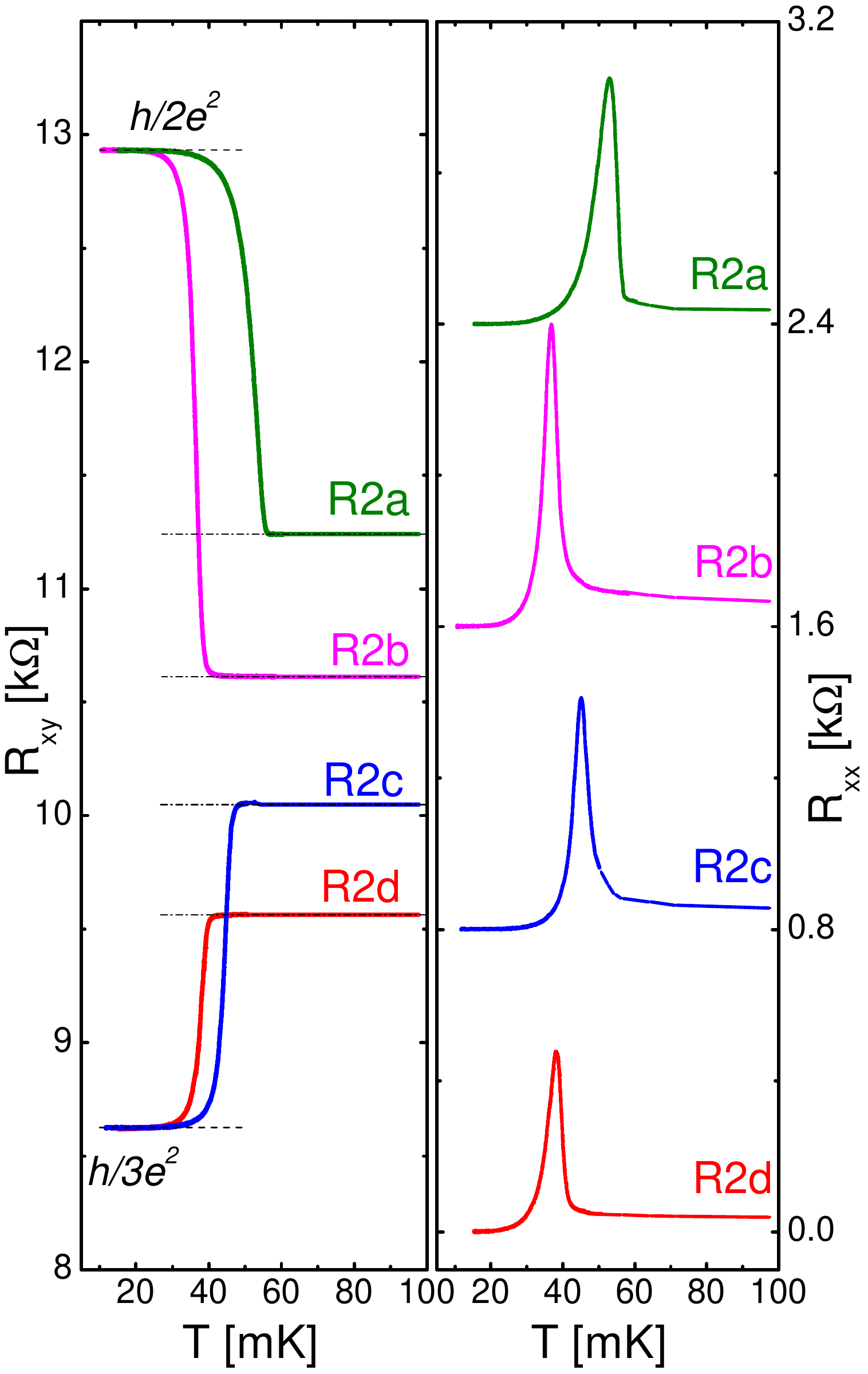}\label{FigC-17}}
}
\end{figure}

\begin{figure}[t]
\centerline{\includegraphics[width=1\columnwidth]{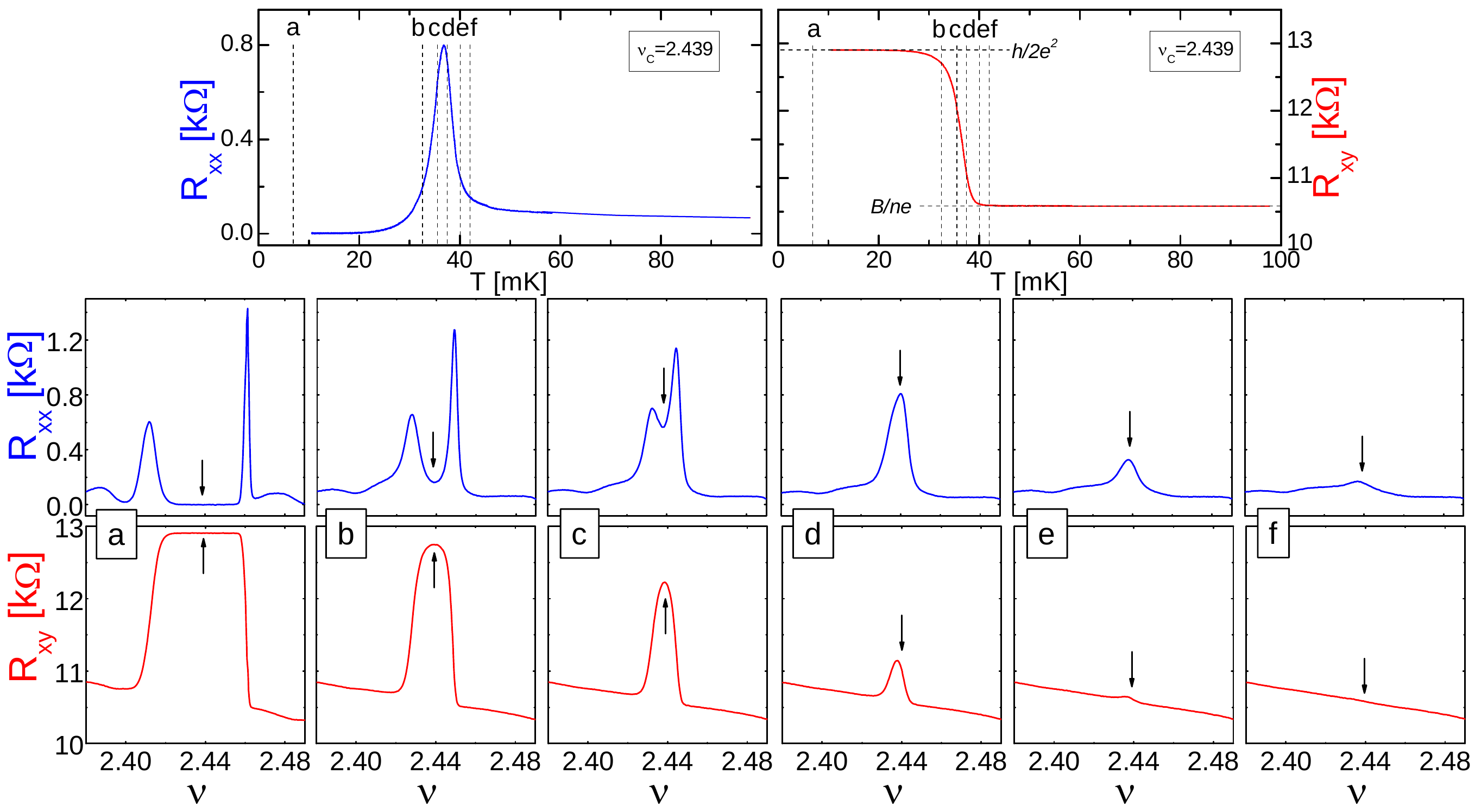}}
\caption{Correlation of cuts along different axes in the $R_{xx}(\nu,T)$  and $R_{xy}(\nu,T)$ 
manifolds in the region of the RIQHS labeled as $R2b$. The two top panels show the temperature 
dependence at $\nu=\nu_c$, whereas the two lower rows of panels show $\nu$ dependence
at fixed temperatures. Data in panels $a$, $b$, $c$, $d$, $e$, and $f$ are collected at the temperatures  $T=6.9, 
32.6, 35.7, 37.7, 40.0,$ and $41.7$~mK. Arrows mark the central filling factor $\nu=\nu_c$.
Precursor of the RIQHS are seen in panels $d$, $e$, and $f$.  Data taken from Ref.\refcite{deng12}.}
\label{FigC-18}
\end{figure}

As the temperature is raised, the two sharp peaks in $R_{xx}$ in the flanks of RIQHSs
persist\cite{deng12}. However,
the two sharp peaks in $R_{xx}$ move closer to each other and, as a result,
the width of the vanishing $R_{xx}$ plateau shrinks.
Such a trend may be seen in Fig.\ref{FigC-16} for several RIQHSs
and may be examined in detail in the middle row of panels of Fig.\ref{FigC-18} 
for one of the RIQHSs labeled  $R2b$. 
At the temperature of $T=32.6$~mK the magnetoresistance shown in Fig.\ref{FigC-18}b  
no longer vanishes,
but it consists of two peaks with a non-zero local minimum in between them.
The location in filling factor of this minimum is
$T$-independent to a good degree and defines the 
central filling factor $\nu_c$ of the RIQHS. 
At $T=35.7$~mK the two spikes of $R_{xx}(\nu)$ have
moved closer to each other; between them there is still
a local minimum, albeit with a large resistance.

A second transport feature associated with the RIQHSs becomes apparent
by measuring the temperature dependence  $R_{xx}(T)|_{\nu}$ at a fixed $\nu$ or $B$-field\cite{deng12}.
One may visualize such a $R_{xx}(T)|_{\nu}$ curve as a
cut in the  $R_{xx}(\nu,T)$ manifold at a given filling factor of choice $\nu$.
Such  $R_{xx}(T)|_{\nu}$ and  $R_{xy}(T)|_{\nu}$ curves measured at $\nu=\nu_c$
are shown in Fig.\ref{FigC-17} for the RIQHSs of the lower spin branch of the second Landau level.
It was found that, as the temperature is lowered,
the Hall resistance $R_{xy}(T)|_{\nu=\nu_c}$ changes from its classical value to
its quantized value, either $h/2e^2$ or $h/3e^2$. At the same time the curve
$R_{xx}(T)|_{\nu=\nu_c}$ exhibits a sharp peak. The inflection point in $R_{xy}(T)|_{\nu=\nu_c}$
and the sharp peak in $R_{xx}(T)|_{\nu=\nu_c}$ coincide and may be interpreted as the
onset temperature of the RIQHS. 
An analysis of the onset temperatures of the RIQHSs in the second Landau level obtained this way
found that they scale with the Coulomb energy, demonstrating the collective nature of these RIQHSs\cite{deng12}.
Fig.\ref{FigC-18} may be used to correlate data collected at constant $\nu=\nu_c$ with data at
constant temperature for both $R_{xx}$ and $R_{xy}$.

\subsection{The precursor of reentrant integer quantum Hall states}

When comparing the $T=35.7$ and $37.7$~mK traces shown in Fig.\ref{FigC-18}c and 
Fig.\ref{FigC-18}d, respectively, one notices that a modest
increase in $T$ of only $2$~mK results in a qualitative change in the magnetoresistance
of a RIQHS from a double peak structure to a single peak in $R_{xx}(\nu)$.
As the temperature is further raised, this single peak rapidly decreases until it merges
into a low resistance background. 
Since the single peaks in $R_{xx}$ marked by arrows 
in Fig.\ref{FigC-18}d, Fig.\ref{FigC-18}e, and Fig.\ref{FigC-18}f
may also be associated with RIQHSs, even though the magnetoresistance does not vanish
and the Hall resistance is far from quantization.
These single peaks are thus signatures of RIQHSs at the highest temperatures and
they may be associated with precursors of the RIQHSs. Simultaneously with the
described changes in $R_{xx}$, $R_{xy}$ evolves from the quantized
value $h/2e^2$ at the lowest temperatures toward its classical value $B/ne$.

Precursors of RIQHSs, as defined above, can be seen for each RIQHS both in the second Landau level\cite{deng12}
(see Fig.\ref{FigC-16}) and also the third Landau level\cite{deng12b}. The concept of precursor of a RIQHS
also provides a natural explanation for the
observation of single peaks in $R_{xx}$ in several experiments
in the regions of the RIQHSs\cite{dean08,choi08}. 
Furthermore, eight precursors of RIQHSs also develop in the second Landau level
of a sample\cite{nodar11} of very low density $n=8.3 \times 10^{10}$~cm$^{-2}$; 
these precursors are marked by yellow shading in Fig.\ref{FigC-8}.

Recently RIQHSs have also been observed in high quality graphene\cite{gr-riqh}. These results highlight
the universal, host-independent physics at play in the formation of the RIQHSs.
It is interesting to note
that in Ref.\refcite{gr-riqh} only the RIQHS labeled $R6a$ is fully developed, i.e. for which $R_{xx}=0$
and Hall resistance quantized to an integer.
At other symmetry related filling factors in the regions labeled $R6b$, $R7a$, and $R7b$,
a peak in $R_{xx}$ was observed which, following the terminology discussed, is the signature
of precursors of RIQHSs.

\subsection{Proliferation of reentrant integer quantum Hall states in the second Landau level}

\begin{figure}[b]
\centerline{\includegraphics[width=1\columnwidth]{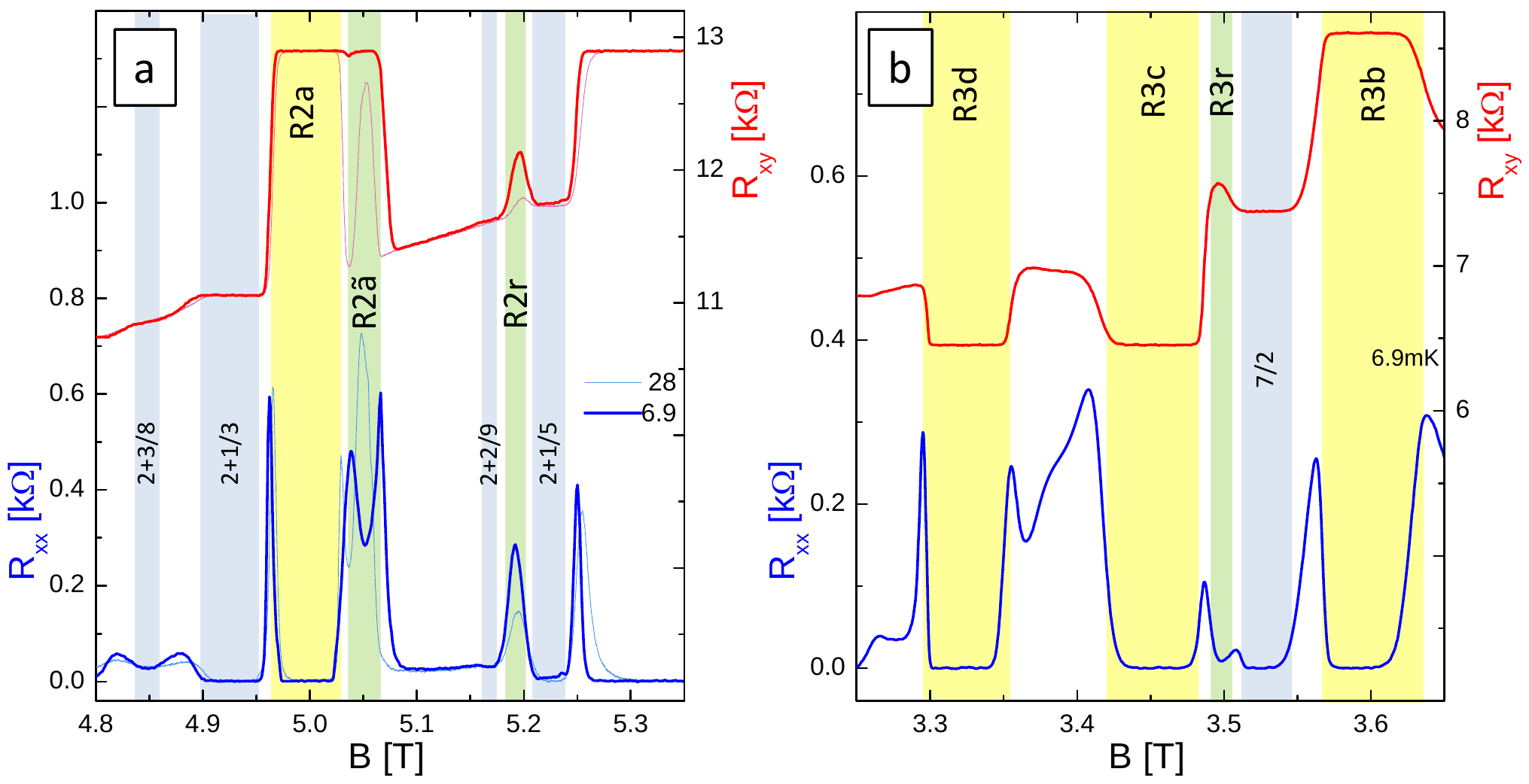}}
\caption{Magnetotransport in the second Landau level exhibiting an increased number of RIQHSs.
RIQHSs discovered in Ref.\refcite{eisen02} are shaded in yellow, any additional RIQHSs
are shaded in green, and FQHSs are shaded in blue.
The $T=6.9$~mK traces in panel $a$ are from Ref.\refcite{deng12b}. 
Traces in panel $b$ are from Ref.\refcite{ethan15}.
}
\label{FigC-19}
\end{figure}

At the time of the discovery of the RIQHSs  in the second Landau level, eight such RIQHSs were observed\cite{eisen02}.
These RIQHSs are some of the most fragile, as their onset temperatures do not exceed $55$~mK\cite{deng12,deng12b}.
Besides these eight RIQHSs, in the second Landau level there are three other magnetoresistance
features that can be associated with additional RIQHSs.

One such additional RIQHSs was discovered as a split-off phase of the RIQHS in the second Landau level
developing at the largest magnetic field\cite{xia04}. This split-off phase, 
labeled $R2\tilde{a}$ in Refs.\refcite{deng12,shingla}, is fully developed, since 
at the lowest temperatures accessed $R_{xx}=0$ and $R_{xy}=h/2e^2$.
This RIQHS is also shown in Fig.\ref{FigC-19}a.

Another developing RIQHS at $2+1/5< \nu <2+2/9$
was identified in Ref.\refcite{deng12b} and it is labeled as $R2r$
in Fig.\ref{FigC-19}a. Using the terminology introduced earlier, the magnetoresistive behavior 
of $R2r$ is consistent with that associated with a precursor of a RIQHS. Indeed, 
the peak in $R_{xx}$ increases and the value of $R_{xy}$ moves closer to $h/2e^2$
as the temperature is lowered. 
The nature of $R2r$ is uncertain, but it may be related to the reentrant insulating phase
of the lowest Landau level that forms in the range between $1/5 < \nu < 2/9$ range
and which was associated with the reentrant Wigner crystal\cite{rip}.
$R2r$ could also be related to the Wigner crystal forming in the flanks of integer plateaus.

In addition, an interesting incipient RIQHS was also detected in the upper spin branch of the second Landau
level, near $\nu=7/2$\cite{ethan15}. This incipient RIQHS is labeled with $R3r$ in Fig.\ref{FigC-19}b.
The magnetoresistance in this region has a distinct local minimum. An intriguing feature of this ground state
is that instead of a Hall resistance moving toward $h/4e^2$, the Hall resistance appears to move towards
$h/3e^2$. This latter property suggests that $R3r$ is different from other RIQHSs nearby, such as the one
labeled $R3c$ in Fig.\ref{FigC-19}b.
 
The development of an increasing number of RIQHs in the second Landau level indicates that the
competition between bubble phases and other ground states in this region is more intricate than
previously thought.

\section{High pressure studies of the second Landau level}
\label{cs-sec6}

Hydrostatic pressure is widely used in probing condensed matter systems, such as superconductors
and correlated electron systems. Pressure decreases the
lattice constant of a crystal. This has profound effects on crystal properties,
since the Bloch wavefunctions supported by the crystal are changed. As a consequence, both the electronic
and the phononic degrees of freedom are affected. Of particular importance is the impact
of high pressure on the band parameters of an electronic system.  Examples of quantities tuned by pressure are
the dielectric constant, effective mass, Land\'e g-factor, band energies, and donor energy levels.   

The effect of hydrostatic pressure on 2DEGs in GaAs/AlGaAs structures is well-documented.
Information of the effective mass, band energies, and pressure dependence of the electron density
is found in the early literature\cite{book}. In addition,
high pressure experiments manipulating the g-factor yielded
significant knowledge on the spin polarization of FQHSs forming in the lowest Landau level\cite{kang}.

More recent use of hydrostatic pressures has contributed to the physics of half-filled
Landau level. This section contains a brief account of these experiments.
It is well known that, depending on the number of filled Landau levels, half-filled single layer 2DEGs 
hosted in GaAs/AlGaAs have three distinct ground states.
In the lowest Landau level,
at $\nu=1/2$ and $3/2$, there is a featureless Fermi sea of composite fermions\cite{duGap}. 
In other Landau levels ordered ground states develop. Indeed, in
the second Landau level, at $\nu=5/2$ and $7/2$, fractional quantum Hall
states were found\cite{willett1,eisen02}. Finally, in high Landau levels, 
at $\nu=9/2$, $11/2$, $13/2$, ..., the ground state
is the quantum Hall nematic\cite{lilly99,du99}. The quantum Hall nematic\cite{fradkin} 
is closely related to the stripe phase predicted by the Hartree-Fock theory\cite{fogler,moessner}.

A peculiar feature of the ordered ground states forming at half filling is that, until recently, 
in clean enough samples and at low enough temperatures only one type of order seemed to develop.
Thus a phase transition between the two different ordered phases, the FQHS and the quantum Hall nematic,
could not be realized in purely perpendicular magnetic fields. This was surprising for two reasons.
First, early theoretical work indicated that a phase transition between a FQHS and the stripe phase
is allowed\cite{haldane-str}. Second, experimental work has clearly indicated that the $\nu=5/2$ FQHS is
close to a nematic phase. Indeed, experiments in tilted magnetic fields have shown that the
isotropic FQHS at $\nu=5/2$ is superseded by an anisotropic nematic phase at relatively
modest tilt angles\cite{tilt1,tilt2}. Nonetheless, during the three decade long history of experimental studies
at $\nu=5/2$, a nematic phase at this filling factor has never been seen in
magnetic fields applied perpendicularly to the 2DEG. As a result, in a large body
of numerical work focused at $\nu=5/2$, the ground state obtained 
 was typically compared to the Pfaffian; the quantum Hall nematic, with a few exceptions\cite{haldane-str,sim6}, 
was not considered a viable ground state at $\nu=5/2$.

A first indication that in the second Landau level order other than the topological order 
of a FQHS may be present came from an experiment on a low density sample\cite{pan14}
in which an incipient anisotropy was reported at $\nu=7/2$. The
observed resistance anisotropy was 2. Subsequent work on 2DEG at high hydrostatic pressures
revealed resistance anisotropy at both $\nu=5/2$\cite{kate1,kate2} and $7/2$\cite{kate3}. 
Furthermore, similarly to
the levels of anisotropy measured at $\nu=9/2$, the anisotry observed in these latter 
experiments\cite{kate1,kate2,kate3} reached extreme values exceeding 1000.

\begin{figure}[t]
\centerline{\includegraphics[width=.9\columnwidth]{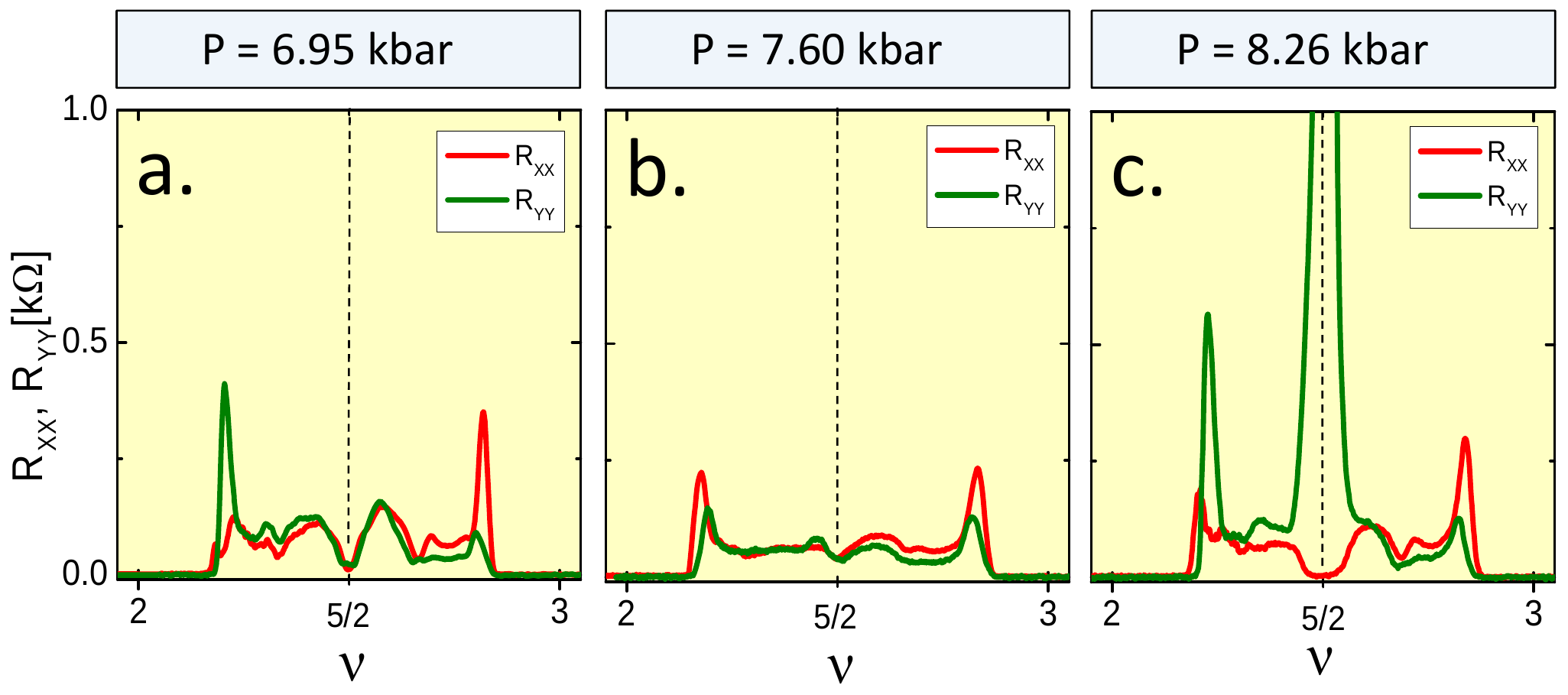}}
\caption{Longitudinal magnetoresistance at $T \approx 12$~mK as measured in 
the second Landau level along two
mutually perpendicular crystallographic axes of the GaAs.
Green traces show $R_{xx}$ measured along the [1$\bar{1}$0] crystal direction, whereas 
red traces show $R_{yy}$ as measured along the [110] crystal direction.
As the pressure is increased, at $\nu=5/2$ the following sequence of ground states is observed: 
an isotropic FQHS (panel a), a nearly isotropic Fermi liquid (panel b), and the quantum Hall nematic (panel c). 
Adapted from Ref.\refcite{kate1}.}
\label{FigC-20}
\end{figure}

In Fig.\ref{FigC-20} we summarize salient features of the longitudinal resistance as measured at three 
different pressures. Since the GaAs/AlGaAs sample was mounted in a commercial
pressure clamp cell, electron thermalization to the base temperature of the refrigerator was no longer 
possible. The electronic temperature in these experiments was estimated to be $T \approx 12$~mK\cite{kate1}. 
Fig.\ref{FigC-20}a, Fig.\ref{FigC-20}b, and Fig.\ref{FigC-20}c show $R_{xx}$ and $R_{yy}$, 
magnetoresistance data collected
along two mutually perpendicular crystal axes of GaAs at the pressures of
$6.95$, $7.60$, and $8.26$~kbar, respectively.
Data shown are consistent with the following sequence of ground states of the 2DEG at
$\nu=5/2$: a rotationally invariant FQHS at $P=6.95$~kbar, 
a ground state close to an isotropic Fermi fluid at  $P=7.60$~kbar, and
a quantum Hall nematic at $P=8.26$~kbar. 
Because at $T \approx 12$~mK the isotropic liquid is observed
in an extremely narrow range of pressures, it was argued that data from Fig.\ref{FigC-20} 
are suggestive of a direct quantum phase transition from the FQHS to the quantum Hall nematic
in the limit of zero temperatures\cite{kate1}. This quantum phase transition occurs 
in the sample studied
at the critical pressure of $P_c \simeq 7.8$~kbar. Since the electron density is pressure dependent,
the density at the critical pressure was $n_c=1.1 \times 10^{11}$~cm$^{-2}$  in this experiment\cite{kate1}.
Later work mapping out the temperature dependence
of the phases strengthened the argument of a direct phase transition in the limit of zero temperatures\cite{kate2}.

There are several reasons to believe that the nematic seen at $\nu=5/2$ at high pressures,
shown in  Fig.\ref{FigC-20}c,
is likely similar in nature to the quantum Hall nematic developing at $\nu=9/2$ and other high value filling factors
in samples in the ambient\cite{kate1}.
First, they are both centered at half-integer filling factors. Second, the temperature dependence
of the magnetoresistance is similar in both cases: it is isotropic above the nematic onset temperature
and it has a very abrupt, almost exponential onset. Third, both the nematic near both $\nu=5/2$
and $\nu=9/2$ develop over a relatively narrow range of filling factors. Indeed, the range
of fillings for the nematic in both cases is about $\Delta \nu \simeq 0.15$. This value is in
sharp contrast to the significantly larger $\Delta \nu \simeq 0.6$ range of filling factors 
of nematicity developing near $\nu=5/2$ in tilted magnetic fields. This is significant, because
an ordered phase induced by an external intensive parameter is not necessarily identical
to a spontaneously ordered phase. For example, the magnetized phase induced in 
a paramagnet when placed in an external magnetic field is not identical to the
spontaneous ferromagnetic phase, as the two do not share the same correlation functions.
In tilted field experiments on 2DEGs the symmetry breaking field favoring nematicity 
is the in-plane component of the magnetic field. While the nematic phase at $\nu=5/2$
forming in tilted magnetic fields is likely related to the nematic forming at $\nu=5/2$ at high pressure
but in the absence of any symmetry breaking fields, the exact nature of this relationship is yet to be determined.

 The importance of data shown in Fig.\ref{FigC-20} is twofold. First, it was established that
the quantum Hall nematic may be stabilized at $\nu=5/2$ in the absence of an in-plane
magnetic field. Second, it was pointed out that the phase transition from a FQHS to the quantum Hall nematic
is of a special type\cite{kate1}. 
Indeed, this phase transition involves a FQHS, which is a topological phase, and the quantum Hall nematic,
which is a transitional Landau phase with a broken symmetry. Phase transitions between two Landau phases are well
known. Furthermore, recent intense investigations of phase transitions between two topologically distinct phases 
have significantly contributed to their understanding. However, phase transitions between the two
distinct classes of phases, such as the transition from a fractional quantum Hall state to the nematic, 
remain rare and present an opportunity for further theoretical development\cite{th1,th2,th3,th4,th5,th6}.
 
One may gain further insight into the transition by investigating it in the upper spin branch
of the second Landau level. Such studies confirmed a qualitatively similar phase transition
at filling factor $\nu=7/2$. It was shown that in a purely perpendicular magnetic field
only even denominator FQHSs may be involved in the FQHS-to-nematic phase transition, highlighting 
therefore an interesting competition between paired states of composite fermions and nematicity\cite{kate3}.
Furthermore, in this study it was also shown that the role of the pressure is to tune the 
electron-electron interaction; data suggest that a FQHS-to-nematic phase transition may also 
be induced by other means of tuning the electron-electron iteration\cite{kate3}.

\section{Conclusions}

In this chapter we addressed a few topics on the electron gas confined to GaAs/AlGaAs hosts.
The study of this system presents an opportunity to learn about the
effects of the Coulomb interaction on the various topological phases and their 
competition with charge ordered phases in a clean environment. 
We discussed recently discovered FQHSs, the effect of the disorder on
the even denominator FQHSs, novel transport features of the RIQHSs,
and recently discovered phase transitions at even denominator filling factors.
Many of these results were enabled by the use of ultra-low temperature 
and of high pressure techniques.
However, the topics covered barely scratched the surface of the ongoing research on this and related systems.
As the purity of different materials improves, the effects of interactions in those materials
will come to the fore. It is then expected that the interplay of
interactions and topology will lead to new physics beyond that of single-electron
band structure in the growing family of topological materials.
Other results, such as the competition of pairing and nematicity in half-filled Landau levels, 
revealed a strong connection between the 2DEG and other strongly correlated materials.
Research on the 2DEG in the fractional quantum Hall regime 
will certainly impact efforts on these materials and
will continue to present future opportunities for discovery.

\section*{Acknowledgments}

I would like to thank Dan Tsui for introducing me to fractional quantum Hall physics
and Jian-Sheng Xia for sharing information about the He-3 immersion cell.
Measurements would have not been possible without the samples of exceptional quality
grown by Loren Pfeiffer and Kenneth West at Princeton and by my colleague at Purdue, Michael Manfra.
I am grateful to Ashwani Kumar, Nodar Samkharadze, Nianpei Deng, Ethan Kleinbaum,
Katherine Schreiber, and Vidhi Shingla for their tireless work in the lab and for their original
contributions to the topics discussed.
Last but not least, I have benefited from numerous discussions with Nicholas d'Ambrumenil, Rudro Biswas, 
Rui-Rui Du, Eduardo Fradkin, James Eisenstein, Lloyd Engel, Jainendra Jain, 
Koji Muraki, Wei Pan, Zlatko Papi\'c, Steve Simon, Jurgen Smet, and Michael Zudov.
This work was supported by the NSF-DMR 1904497 and by the DOE BES award DE-SC0006671.

\bibliographystyle{ws-rv-van}

\end{document}